\documentstyle[12pt,aaspp4,lscape,flushrt]{article}

\def\arcmin{^{\prime}}
\def\arcsec{^{\prime\prime}}

\def\minpoint{'\mskip-7.2mu.\,}
\def\secpoint{''\mskip-7.6mu.\,}
\def\uprime{$U^{\prime}$}
\def\ang{~{\rm \AA}}

\def\cm2{~\rm cm^{-2}}

\def\qno{q_0}

\def\hno{{\rm H}_0}

\begin{document}

\title{Evidence for a Gradual Decline in the Universal Rest-Frame UV
Luminosity Density for $z < 1$}
\author{Lennox L. Cowie\altaffilmark{1,2}, Antoinette Songaila \& Amy J. Barger\altaffilmark{1,2}} 
\affil{Institute for Astronomy, University of Hawaii, 2680 Woodlawn Drive,
  Honolulu, HI 96822\\}

\altaffiltext{1}{Visiting astronomer, W. M. Keck
  Observatory, jointly operated by the California Institute of Technology and
  the University of California.}

\altaffiltext{2}{Visiting Astronomer, Canada-France-Hawaii Telescope, operated 
by the National Research Council of Canada, the Centre National de la 
Recherche Scientifique of France and the University of Hawaii.}

\vskip 1in

\centerline{To be published in {\it Astronomical Journal}, August 1999}

\begin{abstract}

We have utilized various magnitude-limited samples drawn from an extremely
deep and highly complete spectroscopic redshift survey of galaxies observed in
seven colors in the Hawaii Survey Fields and the Hubble Deep Field to
investigate the evolution of the universal rest-frame ultraviolet luminosity
density from $z = 1$\ to the present.  The multi-color data (\uprime , $B$,
$V$, $R$, $I$, $J$, $HK^{\prime}$) enable the sample selection to be made in
the rest-frame ultraviolet for the entire redshift range. Due to the large
sample size and depth ($U_{AB} = 24.75$, $B_{AB} = 24.75$, $I_{AB} = 23.5$),
we are able to accurately determine the luminosity density to $z = 1$.  We do
not confirm the very steep evolution reported by Lilly et al.\ (1996) but
instead find a shallower slope, approximately $(1+z)^{1.5}$\ for $\qno = 0.5$,
which would imply that galaxy formation is continuing smoothly to the present time
rather than peaking at $z = 1$.  Much of the present formation is
taking place in smaller galaxies.  Detailed comparisons with other recent
determinations of the evolution are presented.

\end{abstract}


\section{Introduction} \label{intro}

Over the past several years it has become widely accepted that the global star
formation rate --- at least as seen in the optical and ultraviolet light ---
had a strong peak at around $z = 1$\ and then fell very steeply at lower
redshifts (e.g.\ Pozzetti, Madau \& Dickinson 1998 and references therein).
This would mean that most of the integrated total star formation producing
optically visible light would have occurred at roughly half the present age of
the Universe.  However, the argument for the steep fall-off at $z < 1$\ relies
on a single important analysis by \markcite{lilly96}Lilly et al.\ (1996) based
on the Canada France Redshift Survey (CFRS) sample. This sample has
two weaknesses when it is used to determine the
evolution of rest-frame UV light.  The first of these is that it is a
red ($I$\ band) selected sample with $V$\ and $I$\ photometry, primarily, and
only partial $B$\ and $K$\ coverage.  At low redshifts this requires a very
substantial extrapolation across the $4000\ang$\ break to obtain a $2800\ang$\
rest-frame luminosity.  The second weakness is that the CFRS is slightly too 
shallow for this problem ($I_{AB} = 22.5$, or $I
\sim 22.1$\ in the Kron-Cousins system).  At the highest redshifts, near $z =
1$, the sample does not probe deep enough in the luminosity function to allow
a reliable extrapolation to a total luminosity density.  

In this paper we utilize a large, extremely deep, and highly complete 
spectroscopic redshift survey of galaxies observed in seven colors (\uprime\
[$3400\pm 150\ang$], $B$, $V$, $R$, $I$, $J$, $HK^{\prime}$; the limits in the
blue and red are $B_{AB} = 24.75$\ and $I_{AB} = 23.5$, respectively) in the
Hawaii Survey Fields and the Hubble Deep Field to
investigate the rest-frame UV luminosity density evolution 
from $z = 1$\ to the present. With this sample we are able to avoid 
the above problems encountered in the CFRS study. First,
the availability of both ultraviolet and blue data means that we can {\it
select\/} objects based on their rest-frame ultraviolet magnitudes at all
redshifts and hence avoid the serious problem of selecting at redder
wavelengths and extrapolating to obtain the UV colors.  Second, the
substantial additional depth of our sample allows us to probe to the flat
segments of the luminosity function, thereby giving a more accurate
determination of the luminosity density function. It also enables us to extend
the results to redshifts beyond $z = 1$.  This uniform sample selection
strongly minimises possible errors due to dust in the galaxies, unless there
is substantial evolution of galaxy dust properties with redshift.  Thus, the
shape of the present UV luminosity density evolution should be an accurate
representation of the shape of the star formation history to within the one
residual uncertainty of differential dust evolution.  

We find that the $z < 1$\
UV luminosity density falls with a shallower slope --- roughly
$(1 + z)^{1.5}$\ for $\qno =0.5$\ --- than reported by Lilly et al.\ (1996).
Our low redshift UV luminosity density value is considerably higher than that 
found by Lilly et al.\ and agrees with the local UV luminosity
density recently determined by \markcite{treyer}Treyer et al.\ (1998).
Consquently, much of the integrated total of optically visible stars 
is still being assembled at the present time.

The outline of the paper is as follows.  In \S 2 we provide a brief,
self-contained analysis of the distribution of light with redshift in the
Hubble Deep Field.  If this small field is representative, then the analysis
can be used to demonstrate the primary conclusion of the paper, namely that
the fall-off in the rest-frame ultraviolet luminosity density below $z = 1$\
is relatively shallow.  We then proceed to the main analysis.  We discuss the
data in \S 3; the photometry and the statistical and systematic errors are
presented in an Appendix for textual clarity.  In \S 4 we construct the
rest-frame ultraviolet luminosity functions and the UV luminosity density as a
function of redshift. We then present comparisons with the results of other
analyses.  In order to be fully self-consistent, we specifically restrict
ourselves to the evolution of the ultraviolet luminosity density and do not
attempt to compare with studies of emission-line evolution.  In \S 5 we
briefly summarize our conclusions.

\section{Preliminary Digression} \label{prelim}

The Hubble Deep Field (HDF; \markcite{will96}Williams et al.\ 1996) provides
our deepest view of the faint galaxy populations.  The color information in
the HDF has been extensively used to estimate the high-$z$\ star formation
history of optically-selected galaxies (e.g.\ \markcite{mad97}Madau et al.\
1997; \markcite{con}Connolly et al.\ 1997; \markcite{sawicki}Sawicki, Lin \&
Yee 1997).  As large numbers of spectroscopic
redshifts have accumulated for HDF galaxies (\markcite{steid}Steidel et al.\
1996; \markcite{cohen}Cohen et al.\ 1996; \markcite{low}Lowenthal et al.\
1997; \markcite{phillips}Phillips et al.\ 1997; \markcite{song97}Songaila
1997; \markcite{barg}Barger et al.\ 1999), it is now possible to use these
spectroscopic data to examine the history of star formation at $z < 1$\ and to
show directly that, within the HDF itself, a substantial fraction of the star
formation is occurring at low redshift.  A similar conclusion is reached by
\markcite{pasc98}Pascarelle et al.\ (1998) based on a very different
photometric redshift analysis.

The sky surface brightness of the integrated galaxy light computed from the
\markcite{will96}Williams et al.\ objects in the deeper areas of the WFC chips
--- an area of $4.9~{\rm arcmin}^2$ --- is shown by the solid line in
Figure~\ref{fig:1}.  There are 96 objects in this region of the HDF that now
have spectroscopic identifications. These identified objects, which are
generally the brighter galaxies, contain most of the optical and roughly half
of the $3000\ang$\ light.  We have subdivided the contributions by redshift,
showing the light from known $z < 1$\ galaxies as the solid boxes and that
from known $z < 0.5$\ galaxies as solid diamonds.  The dashed line, which
shows the combined light from all known $z > 1$\ galaxies and {\it all\/}
unidentified or unobserved objects, represents an extreme upper bound to the
fraction of light arising from galaxies at $z > 1$.

Since some (possibly substantial) fraction of the fainter unidentified
galaxies will lie at $z < 1$, the ratio of the $z > 1$\ light to that at lower
redshift is overestimated here but can already be used to make the argument
that a large fraction of star formation is local.  As was pointed out in
\markcite{cowie88}Cowie (1988), \markcite{song90}Songaila et al.\ (1990), 
and \markcite{mad97}Madau et al.\ (1997), the integrated massive star 
formation, which can be characterized by the local metal density, is directly 
proportional to the observed-frame ultraviolet sky brightness, independent of 
cosmology, through the relation
\begin{equation}
\rho_{\rm metals} = {{4\pi\,S_{\nu}} \over {\varepsilon\,c}}
\end{equation}
Estimates of the normalizing factor $\varepsilon$\ can be found in
\markcite{song90}Songaila et al.\ (1990) and \markcite{mad97}Madau et al.\ 
(1997).  Applying this relation to the $3000\ang$\ light in Figure~\ref{fig:1}, 
we see immediately that comparable amounts of metals have been produced by 
$z < 0.5$\ galaxies and by $z > 1$\ galaxies, since the ultraviolet sky 
brightness of the known $z < 0.5$\ galaxies is already 54\% of the maximum 
possible $z > 1$\ light.  
However, in this simple form, the argument depends onthe assumption that the
light from star-forming galaxies is roughly flat in $f_{\nu}$.

A more conservative approach is to compare the 
known $z < 0.5$\ $3000\ang$\ light with the possible $z > 1$\ $6020\ang$\ 
light.  This comparison at equivalent rest-frame wavelengths allows for 
the shapes of the spectral energy distributions (SEDs) and for dust 
extinction.  It also takes into account the disappearance of very 
high redshift galaxies from the $3000\ang$\ band.  The ratio 
$R_{SF} \equiv {\rm light}(z < 0.5)/{\rm light}(z > 1)$\ from 
Figure~\ref{fig:1} is then 26\%.  This
represents a very extreme lower bound on the ratio of the integrated star
formation at $z < 0.5$\ to that at $z > 1$.

The implications for the evolution of the low-$z$\ SFR
[$\dot{\rho}_{\ast}(z)$] can be shown with the simple model
\begin{equation}
 \begin{array}{rcll}
   \dot{\rho}_{\ast}(z) &=& A\,(1 + z)^{\alpha}  & ,\quad z < 1 \nonumber\\ 
                        &=& 2^{\alpha}\, A       & ,\quad z > 1 
 \end{array}
\end{equation}
that can be integrated analytically.  For $\qno = 0.5$\ $R_{SF} = $18\% for
$\alpha = 4$, 30\% for $\alpha = 3$\ and 48\% for $\alpha = 2$.  Thus, even
comparing the {\it candidate\/} $z > 1$\ visual light with the {\it known\/}
$z < 0.5$\ ultraviolet light requires $\alpha \ll 3.2$.  Provided only that
the small HDF is a representative field, even the very conservative and
extremely robust version of the argument requires a shallower slope than has
been inferred from the CFRS data ({\it viz}., $3.9 \pm 0.75$).  The remainder
of the paper verifies this conclusion using our wide-field ultraviolet and
optically-selected samples.

\section{Data Sets} \label{data}

The primary data sets used in the present study are two $6\arcmin \times
2\,\minpoint 5$\ areas crossing the HDF and the Hawaii Survey Field SSA22
(\markcite{lcg}Lilly, Cowie \& Gardner 1991).  In each case the field has been
imaged in 7 colors --- \uprime ($3400 \pm 150\ang$), $B$, $V$, $R$, $I$, $J$\
and $HK^{\prime}$ --- where $R$\ and $I$\ are Kron-Cousins and the
$HK^{\prime}$\ ($1.9 \pm 0.4~\mu$) filter is described in
\markcite{wains}Wainscoat \& Cowie (in preparation).  The images were obtained
on the Keck~II telescope using LRIS (\markcite{oke}Oke et al.\ 1995) and on
the University of Hawaii 2.2m and the Canada-France-Hawaii 3.6m telescopes
using QUIRC (\markcite{hodapp}Hodapp et al.\ 1996), ORBIT, and the UH8K CCD
Mosaic Camera built by Metzger, Luppino, and Miyazaki.  The fields, whose
sizes were set by the usable LRIS area, are fully covered in all colors.  All
magnitudes were measured in $3\arcsec$\ diameter apertures and corrected to
total magnitudes following the procedures in \markcite{cowie94}Cowie et al.\
(1994).  A further two similarly sized areas covering the Hawaii Survey Fields
SSA4 (\markcite{cowie94}Cowie et al.\ 1994) and SSA13 (\markcite{lcg}Lilly,
Cowie \& Gardner 1991) with $B,V,I,HK^{\prime}$ data were used to augment the
SSA22 and HDF $B$, $I$, and $HK^{\prime}$-selected samples.  The fields,
areas, magnitude limits, and numbers of spectroscopically-identified galaxies,
stars, and unknowns (unobserved or unidentified objects) for each of the
color-selected samples are given in Table~\ref{tbl:1}. 
A more detailed discussion and analysis of
the photometric and spectroscopic samples used in this paper can be found in
Barger et al.\ (1999).

\section{Rest-Frame Ultraviolet Luminosity Density Evolution} \label{uvlum}

\subsection{Construction of the luminosity functions}

For redshifts, $z$, for which the rest-frame UV wavelength, $\lambda_U$, maps
directly to the wavelength of one of the observed color bands, the absolute
rest-frame UV magnitude, $M_U$, in the $AB$\ system ($M_U = 0$\ corresponds
to $\log_{10}\,F_{\nu} = -48.57$) is given by
\begin{equation}
M_U = m_{AB} + 2.5\,\log_{10}(1+z) - 5\,\log_{10}(d_L(z)/10\,{\rm pc})
\end{equation}
where $m_{AB}$\ is the observed magnitude at the redshifted wavelength
$\lambda_U(1+z)$\ in the $AB$\ system and $d_L(z)$\ is the luminosity
distance.  For redshifts such that $\lambda_U(1+z)$\ differs by a small amount
from this central wavelength, we can apply the same formula
but with the addition of a small differential $k$\ correction $dk(z)$, 
defined as
\begin{equation} 
dk(z) = -2.5\,\log_{10}\,{{f_{\nu}\,(\lambda_U(1+z))} \over
{f_{\nu}\,(\lambda_U(1+z_C))} }
\end{equation}
where $f_{\nu}$\ is the SED of the galaxy, and $z_C$\ is the redshift
corresponding to the center of the band.  Because of the frequent,
regularly-spaced color sampling, we can simply construct $dk(z)$\ by
interpolation from the neighboring color bands.  The value of $dk(z)$\ is
generally small (less than 0.2~mag) for the redshift intervals which are used.

The $2500\ang$\ rest-frame absolute magnitudes, computed in the appropriate 
redshift intervals from the \uprime, $B$, $V$\ and $I$\ samples, 
are shown versus redshift in Figure~\ref{fig:2}.  In the overlapping redshift
intervals the absolute magnitudes compare well.  The rise in the maximum
absolute UV rest-frame luminosity with redshift mapped by the upper envelope
of the distribution is partly a reflection of the larger volume sampled at
high redshift but also partly reflects the known evolution that causes the
maximum UV luminosity to increase with redshift (\markcite{lilly96}Lilly et
al.\ 1996; \markcite{cowie94}Cowie et al.\ 1996); we will
discuss this in more detail below.

We adopt the traditional $V_{max}$\ method of \markcite{felten}Felten (1977)
for constructing the luminosity functions.  The number density of galaxies in
the redshift range [$z_1,z_2$] with magnitude $M$\ is given by
\begin{equation}
\phi(M)\,dM \propto \sum\, {{1} \over {V_{max}(M)}}
\end{equation}
where the sum is over all galaxies with magnitude $M \pm dM/2$.  $V_{max}(M)$\
is the maximum total volume in all the samples where galaxies with absolute
magnitude $M$ are observable in the appropriate apparent magnitude range and
lie in the redshift range.  (A very complete description of the procedure may
be found in \markcite{ellis96}Ellis et al.\ 1996). The ultraviolet luminosity
density, $\ell$, is then
\begin{equation}
\ell = 4.4 \times 10^{20}\, \sum {{10^{-0.4M}} \over {V_{Max}(M)}}\quad {\rm
ergs\ s^{-1}\ Hz^{-1}}
\end{equation}
where the sum is over all the observed objects.  We use Schechter function
fits with $\alpha = -1$\ to extend $\ell$\ from the brighter absolute
magnitudes observed to $M = -16$, which we then use as our total luminosity
density.

In the computation of the luminosity functions, we assign errors based on the
Poisson distribution corresponding to the number of galaxies contributing to
the absolute magnitude bin.  However, a larger potential source of error is
the missing or unidentified galaxies in each sample.  Following the standard
procedure, we compute the luminosity function after assuming that these
objects follow the redshift distribution of the identified objects, but
because of systematic effects, this assumption may not be valid.  We have
therefore also computed the LFs with all, and also with none, of the missing
objects allocated into the redshift bin, which should provide extremal
estimates of this error.  In the case where we allocated the missing objects
into the redshift bin, the allocated redshifts were randomly distributed
uniformly within the redshift interval.  We are able to use this very robust
procedure because of the extremely high completeness of the spectroscopic samples.

\subsection{UV luminosity functions versus rest wavelength}

A key issue is the optimum wavelength at which to compute the rest-frame LFs
and the UV light densities.  Previous efforts, such as those by the CFRS, have
computed the light densities at a relatively long wavelength of $2800\ang$,
which is optimal for a red sample and minimises dust extinction but 
may be a poorer measure of massive star formation rates than shorter
wavelengths.  The local UV sample of \markcite{treyer}Treyer et al.\ (1998),
which provides an invaluable low redshift comparison, is at $2000\ang$.
However, our own sample gives the best combination of wide redshift range and
largest galaxy sample at a rest wavelength around $2500\ang$.

In the following discussion we shall use all three of these rest wavelengths
for specific comparisons and analyses. Before proceeding, however, we use the
various color samples to show that at $z = 1$\ the luminosity functions and
light densities are only weakly sensitive to the wavelength choice.  At $z =
1$\ the \uprime, $B$\ and $V$\ samples correspond to rest-wavelengths of
$1700\ang$, $2250\ang$\ and $2750\ang$. In Figure~\ref{fig:3} we compare the
luminosity functions constructed at each of these rest wavelengths in the
redshift interval $0.7 < z < 1.3$.  It is apparent that the luminosity
functions are quite similar, with only a slight fading of $M_U$\ with
decreasing wavelength.  The corresponding luminosity densities for galaxies
with luminosities greater than $M_U = -17.5$\ (squares), or extrapolated to $M
= -16$\ assuming an $\alpha = -1$\ Schechter function (dashed line), are shown
in Figure~\ref{fig:4}.  The total UV luminosity density above $M = -16$\ in
this redshift interval is fit by a power law
\begin{equation}
\ell = 2.6\times 10^{26}\,h_{65}\,\left( {{\lambda} \over
{3000~\rm\AA}}\right)^{1.1}\,\quad {\rm ergs\ Mpc^{-3}\ Hz^{-1}}
\end{equation}
where $h_{65}$\ is the Hubble constant in units of $65~{\rm km\ s^{-1}\
Mpc^{-1}}$.  We shall generally avoid comparing light densities at different
wavelengths.  Where necessary, however, we use the weak wavelength dependence
of equation\ (7), assuming, somewhat arbitrarily, that this is valid for other
redshifts.

\subsection{The redshift evolution of the $2500\ang$\ rest-frame luminosity 
density}

We first construct the $2500\ang$\ rest-frame luminosity function in
three redshift bands: $z = 0.20 - 0.50$\ using the \uprime-selected sample
(Figure~\ref{fig:5}), $z = 0.6 - 1.0$\ using the $B$-selected sample
(Figure~\ref{fig:6}), and $z = 1.0 - 1.5$\ using the $V$-selected sample
(Figure~\ref{fig:7}).  In each case the solid curve is the
incompleteness-corrected luminosity function with $\pm 1~\sigma$\ Poisson
error bars on the symbols; the dotted curve is the luminosity
function obtained from only the observed objects (the minimal
function); and the dashed curve is the luminosity function when all the
unidentified objects are taken to lie within the redshift interval 
(the maximal function).

We next compute the luminosity density for galaxies more luminous than $M_U =
-16$. This is a directly determined quantity in the lowest redshift interval,
but for the higher redshift intervals it requires an extrapolation
which was made with an $\alpha = -1$\ Schechter function fit.
For $z = 0.6 - 1.0$ the observed light density for $M_{AB} \le
-17.5$\ is $1.7\times 10^{26}\,h_{65}\ {\rm ergs\ s^{-1}\ Hz^{-1}\ Mpc^{-3}}$,
and the extrapolated light density is 21\% higher.  For $z = 1.0 - 1.5$\ the
observed light density for $M_{AB} \le -18.25$\ is $1.4\times 10^{26}\,h_{65}\
{\rm ergs\ s^{-1}\ Hz^{-1}\ Mpc^{-3}}$, and the extrapolated light density is
34\% higher.  

The incompleteness-corrected $M_{AB} \le -16$\ light density as a
function of redshift is shown in Figure~\ref{fig:8} as the filled squares;
the redshift range is shown as the thin horizontal lines.  The statistical
and systematic errors (see Appendix) are shown as the thicker portions of the
error bar.  We have performed similar calculations for both the minimal
and maximal functions.  These represent the extremal range of the possible
light density at a given redshift and are shown as the open squares joined by
the vertical thin lines.  The solid line shows a simple 
$(1 + z)^{1.5}$\ evolution law which would result in equal amounts of star
formation per unit redshift interval and would constitute an acceptable fit
to the data.  The dashed line shows a redshift evolution of
$(1 + z)^4$\ matched to the second data point;  this is clearly too steep,
even allowing for the maximum incompleteness correction.

The shape and normalization of the rest-frame UV luminosity function is of
course a consequence of a complex mix of the evolution of the star formation
rates in galaxies of various types and masses and the decrease in $M_{U\ast}$\
with declining redshift, which most likely reflects a preferential drop in the
star formation rates of the most massive galaxies (\markcite{cowie96}Cowie et
al.\ 1996).  Nevertheless, it is interesting to compare the functions to
determine if there is evidence for a change of shape (Figure~\ref{fig:9}).
For $\qno = 0.5$\ the functions can be quite well overlayed with a simple
shift in the absolute magnitude (pure luminosity evolution).  Though there is
a hint that the higher redshift functions might be steeper, there is no
statistically significant difference in the shapes as a function of redshift.
The absolute magnitude shifts used in constructing Figure~\ref{fig:9}
(+0.65~mag at $z = 0.6 - 1.0$\ and +0.75~mag at $z = 1.0 - 1.5$, both relative
to $z = 0.2 - 0.5$) can also be used to check the $2500\ang$\ rest-frame
luminosity density evolution and would imply a ratio of 1 : 1.8 : 2.0 in these
redshift bins; the best-fit power law corresponds to $\ell \propto (1 +
z)^{1.4}$, which is quite similar to the estimates given above.

\subsection{Comparison with other samples}

The most directly comparable sample is the low-redshift, rest-frame
ultraviolet ($2000\ang$) selected sample of Treyer et al.\markcite{treyer}
(1998). To make a direct comparison with this sample we have computed the
$2000\ang$\ luminosity functions at $z = 0.5 - 0.9$\ using the
\uprime-selected sample and at $z = 1.0 - 1.5$\ using the $B$-selected sample.
We compare these with the luminosity function determined by Treyer et al.\ in
Figures~\ref{fig:10} and \ref{fig:11}.  The solid line with the symbols shows
the presently-determined luminosity function. The dashed line shows the Treyer
et al.\ function for $\hno = 65~{\rm km\ s^{-1}\ Mpc^{-1}}$\ and converting
the magnitude system used by Treyer et al.\ to $AB$\ magnitudes with a
2.29~mag offset.  Here we have used a renormalization rather than a magnitude
offset to overlay the functions.  For $z = 0.5 - 0.9$\ the normalization of
the Treyer et al.\ function is multiplied by 1.7. For $z = 1.0 - 1.5$\ the
normalization is multiplied by 2.8.  The corresponding luminosity density
changes would be an increase of $\ell \propto (1+z)^{1.3}$.

The measured $2000\ang$\ luminosity densities in the Treyer et al.\ sample are
compared with the incompleteness-corrected $2000\ang$\ luminosity densities
from this work in Figure~\ref{fig:12}, where once again we have shown the
maximal and minimal luminosity densities with open squares.  As was the case
for the $2500\ang$\ data, the combined data sets exhibit a slow evolution, as
shown by the $(1 + z)^{1.5}$\ solid line.  A steeper dependence, such as the
$(1 + z)^4$\ evolution of the dashed line, is radically inconsistent with the
data.  Because of the steep rise in the Treyer et al.\ luminosity function at the
faintest magnitudes, it is best fit by an $\alpha = -1.6$\ power law rather
than the $\alpha = -1.0$\ power law 
(characteristic of the optical luminosity functions) which we have adopted in
extrapolating the luminosity functions to the fainter magnitudes.  In order to
investigate the dependence on $\alpha$\ we have fitted the $2000\ang$\
luminosity functions at $z = 0.5 - 0.9$\ and $z = 1 - 1.5$\ with Schechter
functions with indices $\alpha = -1.0$\ and $\alpha = -1.5$.  The fitted
parameters and the luminosity densities above $-16$\ are summarized in
Table~\ref{tbl:2} where they are compared with the values derived by Treyer et
al.  Increasing the index from $-1.0$\ to $-1.5$\ has the effect of
preferentially increasing the luminosity density at the higher redshifts where
the extrapolation is larger.  However, the effect is not large.  Using the
$-1.5$\ derived luminosity density would increase the slope to $(1+z)^{1.7}$\
from $(1+z)^{1.3}$\ for the $\alpha = -1.0$\ case.

We next compare the present sample with the \markcite{lilly96}Lilly et al.\
(1996) analysis of the CFRS data at a rest-frame wavelength of $2800\ang$.  In
order to make the most direct comparison possible, we followed the Lilly et
al.\ redshift intervals of [0.2,0.5], [0.5,0.75], [0.75,1], constructing the
luminosity functions with the \uprime\ sample, the $B$\ sample, and the $V$\
sample, respectively.  The luminosity densities constructed from these samples
are shown in Figure~\ref{fig:13}, where the filled squares are again the
incompleteness-corrected luminosity density, the thicker portions of the error
bars show the systematic and statistical errors, and the open squares show the
minimal and maximal light densities.  The open diamonds show the Lilly et al.\
analysis; their three higher redshift points are based on the CFRS data and
the lowest redshift point is based on the data of \markcite{love}Loveday et
al.\ (1992).  The lowest redshift point is already contradicted by Treyer et
al.'s (1998) analysis, shown as the filled diamond, corrected upwards by a
factor of 1.5 for the wavelength difference between $2000\ang$\ and
$2800\ang$.  The Treyer et al.\ point is also comparable with the [0.2,0.5]
Lilly point.  The present data exhibit a much shallower slope than the Lilly
et al.\ data in the $z = 0.2 - 1$\ range.  This results primarily from the
[0.75,1] point being about a factor of two lower than the highest redshift
Lilly point.  The two intermediate redshift points agree well with the CFRS
analysis.

We also compare with the analyses of \markcite{con}Connolly et al.\ (1997) and
\markcite{sawicki}Sawicki et al.\ (1997), shown as open triangles and inverted
open triangles, respectively, in
Figure~\ref{fig:13}. These two analyses are based on photometrically estimated 
redshifts in the small HDF proper and are in broad general agreement with each 
other and with the work of \markcite{pasc98}Pascarelle et al.\ (1998). 
(We do not include a direct comparison to the latter work here 
as it is computed at $1500\ang$.) Following the
discussion in the Appendix, we have made a small 20\% downward correction to
allow for the slightly higher counts in the relevant magnitude range in the
HDF proper versus the wide surrounding field.  Both the [0.5,1] and [1.0,1.5] 
points from these analyses are higher than our incompleteness-corrected
estimate but are consistent within the statistical and incompleteness errors.
Our best estimate value may indeed be slightly low here, since there may be
preferential incompleteness in these redshift ranges.  However, there is also
relatively little validation of photometrically estimated redshifts in the
range $z = 1$\ to 2 where the observed spectral range ($\lambda = 3000\ang$\
to $22,000\ang$) is quite featureless for blue galaxies.  Thus, the high
redshift Connolly et al.\ and Sawicki et al.\ points could be overestimated 
if lower redshift blue
irregulars have been assigned to the wrong bin. Since our maximal
incompleteness-corrected estimate is not substantially different from the
photometrically estimated values, the discussion of \S 5 does not depend on
the resolution of this issue.

\subsection{The relative evolution of the rest-frame UV, blue and red light}

As has been known for several years now (Lilly et al.\markcite{lilly96} 1996;
Cowie et al.\markcite{cowie96} 1996) the rest-frame red luminosity has an
extremely weak evolution with redshift out to $z = 1$.  This slow evolution
can be clearly seen in Figure~\ref{fig:14}, in which we compare the rest-frame
$8000\ang$\ luminosity at $z = 0.2 - 0.5$, obtained using the $I$-selected
sample, with that at $z = 0.7 - 1.3$, computed using the $K$-selected sample.
The two functions can be brought into full consistency with a very small
(0.2~mag) pure luminosity shift corresponding to a decrease of only 20\% in
the red light from $z = 0.8$\ to $z = 0.35$, thereby confirming with the
present sample that the downward evolution of the UV light density by a factor
of two between $z = 1$\ and $z = 0.35$\ does take place against an essentially
invariant shape and normalization for the red light density.

\section{Summary}

We have investigated the evolution of the universal ultraviolet luminosity
density from $z=1$ to the present using a magnitude-limited sample selected on
the basis of rest-frame ultraviolet colors at all redshifts from an extremely
deep and highly complete spectroscopic redshift survey.  Our uniform selection
procedure avoids the serious problem of selecting the sample at redder
wavelengths and then extrapolating to obtain ultraviolet colors, and it
strongly minimises possible errors due to dust in the galaxies, unless there
is substantial evolution of galaxy dust properties with redshift.  The depth
of the current sample is also sufficiently deep to probe to the flat segments
of the luminosity function and to extend the results to redshifts beyond
$z=1$, thereby enabling us to make an accurate determination of the luminosity
density function.

We find that our incompleteness-corrected rest-frame 2500$\ang$ luminosity
densities for $M_U\le -16$ as a function of redshift are well fit by an
$l\propto (1+z)^{1.5}$ evolution law for $\qno = 0.5$.  (The slope would be
shallower for open geometries.) The decline in $M_U$ with decreasing
redshift most likely reflects a preferential drop in the star formation rates
of the most massive galaxies. A direct comparison of the measured low-redshift
2000$\ang$ luminosity density of Treyer et al.\ (1998) with
incompleteness-corrected 2000$\ang$ luminosity densities computed from our
sample also shows that the slow evolution law of $l\propto (1+z)^{1.5}$\
provides a good fit.  When we compare the Lilly et al.\ (1996) CFRS analysis
with our sample analyzed at rest-frame 2800$\ang$, we find that our sample
gives a much shallower slope in the $z=0.2-1$ range.  This most probably
arises in part due to the relatively large extrapolation at low redshifts
Lilly et al.\ needed to make to go from a primarily $V$ and $I$-based data
sample to ultraviolet colors.  The two lowest-redshift Lilly et al.\ points
are also in disagreement with the Treyer et al.\ analysis. Interestingly,
however, the present UV luminosity density analysis leads to a closer
agreement with star formation rates found in an H$\alpha$ analysis of the CFRS
data by Tresse \& Maddox (1998).

We can summarise our results in the form of the widely-discussed plot of UV
luminosity density versus redshift, our version of which is shown for the
$2500\ang$\ rest frame in Figure~\ref{fig:8} and at $2000\ang$\ in
Figure~\ref{fig:12}.  This plot is rather different from the now conventional
`Madau' form which has a strong rise to $z = 1$.  Rather, we see a fairly slow
decline in $\dot\rho_{UV}$\ from $z = 1.5$\ to $z = 0$, which is reasonably
well described by a simple $\dot\rho_{UV} = 7\times 10^{25}\,(1 +
z)^{1.5}~{\rm ergs\ s^{-1}\ Mpc^{-3}\ Hz^{-1}}$\
relation. This result has fairly profound philosophical implications in that
the integrated star formation is continuing to rise smoothly at the present
time and the bulk of the star formation has occurred at recent times.  (This
result holds irrespective of cosmological geometry.)  Put succinctly, as seen
in the optical, now may be the epoch of galaxy formation and not $z = 1$.

\appendix

\section{Systematic and Statistical Errors and Comparison with HDF
Photometry and Number Counts}

Because much of our information on the $z > 1$\ rest-frame UV
luminosity density is based on the small Hubble Deep Field, it is important to
check the consistency of the photometry and number counts between the present
data and the HDF tables of Williams et al.\markcite{will96} (1996).

Since the wide-field area around the HDF fully overlaps the HDF proper, 
we can compare the photometric systems directly.  We first identified the
corresponding objects in the two tables whose centroids lie within 
$0\,\secpoint 4$\ of each other and are not complex objects.  
We then compared our corrected
aperture magnitudes with the total magnitudes in Williams et al.\  The color
equations for the four bands are found to be,
\begin{equation}
 \begin{array}{rcrcl}
   I &=& -0.39 &+& AB_{814} - 0.09\,(AB_{602} - AB_{814}) \\
   V &=&  0.16 &+& AB_{602} + 0.40\,(AB_{602} - AB_{814}) \\
   V &=&  0.14 &+& AB_{602} + 0.47\,(AB_{450} - AB_{602}) \\
   B &=&  0.05 &+& AB_{450}\phantom{ + 0.47\,(AB_{450} - AB_{602})\ \ } \\
   U &=&  0.10 &+& AB_{300} - 0.25\,(AB_{300} - AB_{450}) \\
 \end{array}
\end{equation}
where we have forced the color term in the $B$\ equation to zero.  Allowing
for the offsets in zero points between the color systems ($I - I_{AB} =
-0.33,\ B - B_{AB} = 0.16$), we can see that there is a maximum deviation of
0.15~mag in the $V$-band versus $AB_{602}$, where the color term is, however,
large.  We take this to be the maximum systematic uncertainty in the
photometry.

Using these photometric transformations we next constructed the $V$-band
number counts (galaxies and stars) in the WFPC areas of the HDF itself ($4.9\
{\rm arcmin}^2$) and in the $81\ {\rm arcmin}^2$\ area of the HDF flanking
field region.  This comparison is shown in Figure~\ref{fig:A1}.  There is
broad general agreement in the counts within the small number fluctuations.
Over the range, $21 < V < 24.25$, the HDF proper has a 17\% higher total flux
per unit area than the large flanking field area.  In Figure~\ref{fig:A2} we
show the HDF galaxy counts in $B$\ compared with the average $B$\ galaxy
counts in the regions of the HDF and SSA22 used in the present 
analysis.  (Only spectroscopically confirmed stars have been removed from both
counts.)  The total flux per unit area of objects in the HDF with $21 < B <
24.75$\ is 13\% higher than in the average of the two fields.

Finally, in Figure~\ref{fig:A3} we show a comparison of the total HDF $0 < z <
0.5$\ and $0 < z < 1$\ HDF sky surface brightness in the ranges $21 < V <
24.5$, $21 < B < 24.75$\ and $21 < U^{\prime} < 24.5$\ with the present
samples.  In the analysis the HDF lies about 25\% higher in the total light in
this range, with a disproportionately high amount of $z < 0.5$\ light, which
is between 25\% and 50\% higher in the various colors.  The results emphasize
that clustering can cause fluctuations at this level in fields of the size of
the HDF, and they emphasize the importance of using multiple large areas.  In
order to compare with the HDF-derived data of Connolly et al.\markcite{con}
(1997) we have therefore revised their fluxes downward by 20\%.  However, the
systematic error in the present much larger data sample should be considerably
smaller than this and is dominated by the photometric errors.  We will adopt a
conservative systematic error limit of $\pm 20$\% of the rest-frame UV
luminosity densities. 

In order to obtain a realistic estimate of the statistical error we 
divided the $B$\ sample into two parts --- one comprising the HDF and SSA22
data and one comprising the SSA13 and SSA4 data --- and constructed the
rest-frame $2500\ang$\ luminosity function in the $0.6 < z < 1$\ range from each
subsample.  The two luminosity
functions are shown in Figure~\ref{fig:A4}. The luminosity
densities inferred from the two functions differ by 17\%.  Based on this
comparison, we have assigned a statistical error of $0.2\,(N/40)^{-0.5}$,
where $N$\ is the total number of galaxies in the sample bin, and we have
added this error in quadrature to the systematic error to determine the formal
error bars in the rest-frame UV luminosity density plot.

\acknowledgments


\begin{deluxetable}{ccclccc}
\tablenum{1}
\tablecaption{Color-selected Samples \label{tbl:1}}
\tablehead{
\colhead{Color} & \colhead{Field} & \colhead{Area (${\rm arcmin}^2$)} & \colhead
{Limit}
& \colhead{Galaxies} & \colhead{Stars} & \colhead{Unknowns} \nl
}
\startdata
$U^\prime$ & SSA22 & 15.7 & $U^{\prime} < 24.75$ & 115 & 12 & 6 \nl
& HDF & $17.6$ & $U^{\prime} < 24.5$ & 103 & 10 & 29 \nl
\tablevspace{20pt}
$B$ & SSA22 & $16.0$ & $B < 24.75$ & 166 & 35 & 14 \nl
& HDF & $13.6$ & $B < 24.25$ & 109 & 14 & 25 \nl
& SSA13 & $12.7$ & $B < 24.25$ & 75 & 6 & 7 \nl
\tablevspace{20pt}
$V$ & SSA22 & $15.7$ & $V < 24.5$ & 172 & 43 & 15 \nl
& HDF & $14.3$ & $V < 23.75$ & 87 & 15 & 6 \nl
\tablevspace{20pt}
$I$ & SSA22 & $15.7$ & $I < 23.25$ & 196 & 50 & 25 \nl
& SSA13 & 12.8 & $I < 22.5$ & 95 & 14 & 5 \nl
\tablevspace{20pt}
$K$ & SSA22 & $15.7$ & $K < 20.5$ & 109 & 37 & 24 \nl
& HDF & 14.3 & $K < 20.5$ & 96 & 15 & 29 \nl
& SSA13 & 12.8 & $K < 20$ & 83 & 12 & 22 \nl
& SSA4 & 14.9 & $K < 19$ & 49 & 25 & 17
\enddata
\end{deluxetable}

\clearpage

\begin{deluxetable}{crccc}
\tablenum{2}
\tablecaption{Fits to $2000~\rm\AA$\ Luminosity Functions \label{tbl:2}}
\tablehead{
\colhead{$z$\ Range} & \colhead{$\ \ \alpha$} & \colhead{$\ \ M_{\ast}$\tablenotemark{a}} & \colhead
{$\phi_{\ast}$} & \colhead{$L$} \\
[0.5ex] &&& ($\times 10^{-3}~{\rm Mpc}^{-3}$) & ($\times 10^{26}\ {\rm ergs\ cm^{-2}\ s^{-1}\
Hz^{-1}})$
}
\startdata
1.0 -- 1.5 & $-1.0$ & $-19.1 \pm 0.25$ & 10.0 & 1.8 \nl
           & $-1.5$ & $-19.3 \pm 0.15$ & 0.8  & 2.5 \nl
\tablevspace{0.1in}
0.5 -- 0.9 & $-1.0$ & $-18.2 \pm 0.3$  & 16.0 & 1.2 \nl
           & $-1.5$ & $-18.5 \pm 0.4$  & 11.0 & 1.4 \nl
\tablevspace{0.1in}
Local\tablenotemark{b} & $-1.6$ & $-19.16^{+0.31}_{-0.29}$ & 2.5 & 0.7
\enddata
\tablenotetext{a}{Errors in $M_{\ast}$\ are 98\% confidence limits except for
the local value, for which errors are $1~\sigma$.  All values are calculated
for $H_0 = 65~{\rm km\ s^{-1}\ Mpc^{-1}}$\ and $q_0 = 0.5$.}
\tablenotetext{b}{Treyer et al.\ (1998)}
\end{deluxetable}

\clearpage

\newpage

\figcaption[]{\label{fig:1}Sky surface brightness of the integrated galaxy
light in the Hubble Deep Field.  The solid line shows the integrated light
from all the galaxies, the filled squares that of known $z < 1$\ galaxies, and
the filled diamonds that of $z < 0.5$\ galaxies.  The dashed line shows the
light from galaxies known to be at $z > 1$\ plus all unidentified galaxies.
}

\figcaption[]{\label{fig:2}Absolute $2500~{\rm\AA}$\ rest-frame magnitudes
versus redshift, computed from the \uprime\ (filled squares), $B$\ (open
diamonds), $V$\ (stars), and $I$\ (filled triangles) samples.
}

\figcaption[]{\label{fig:3}Luminosity functions in the redshift interval $0.7
< z < 1.3$\ constructed at rest wavelengths of $1700~{\rm\AA}$\ (open
diamonds), $2250~{\rm \AA}$\ (filled squares), and $2750~{\rm \AA}$\ (open
triangles) from the \uprime, $B$, and $V$\ samples, respectively.
}

\figcaption[]{\label{fig:4}Luminosity density in the redshift interval $0.7 <
z < 1.3$\ computed from the luminosity functions of Fig.~3.  The squares are
the luminosity density for galaxies with luminosity $M_{AB} > -17.5$.  The thin
dashed line represents a power law fit to the data.  The solid line is
$2.6\times 10^{26}\,(\lambda/3000~{\rm\AA})^{1.1}~{\rm ergs\ cm^{-2}\ s^{-1}\
Hz^{-1}}$, which represents the power law fit to the luminosity densities
extrapolated to $M_{AB} = -16$, assuming an $\alpha = -1$\ Schechter function.
}

\figcaption[]{\label{fig:5}$2500~{\rm\AA}$\ rest-frame luminosity function
(diamonds and solid line) in the redshift interval $0.2 < z < 0.5$\
constructed from the \uprime\ sample.  $\pm 1~\sigma$\ error bars are shown.
The dotted line is the luminosity function computed from only the observed
objects (the ``minimal'' function), whereas the dashed line is computed after
placing all of the unidentified objects into the redshift bin (the ``maximal''
function).  
}

\figcaption[]{\label{fig:6}As in Fig.~5 but for the redshift interval 
$0.6 < z < 1.0$, constructed from the $B$\ sample.
}

\figcaption[]{\label{fig:7}As in Fig.~5 but for the redshift interval 
$1.0 < z < 1.5$, constructed from the $V$\ sample.
}

\figcaption[]{\label{fig:8}$M_U < -16$\ light density at $2500\ang$\ as a
function of redshift: directly determined in the interval $0.2 < z < 0.5$\ and
incompleteness-corrected via an $\alpha = -1$\ Schechter function for the
intervals $0.6 < z < 1.0$\ and $1.0 < z < 1.5$.  In each redshift interval the
light density is shown as a filled square and the redshift range as a thin
horizontal line.  Statistical and systematic errors (Appendix) are shown as
the thick portions of the error bar.  Light densities computed using the
minimal and maximal functions (see text and Figs. 5 -- 7) are shown as open
squares joined by thin vertical lines.  The solid line is a $(1 + z)^{1.5}$\
evolution law.  The dashed line shows an evolution of $(1 + z)^4$, normalized
to the second data point.  }

\figcaption[]{\label{fig:9}$2500~{\rm\AA}$\ rest-frame luminosity functions
($\qno = 0.05$)
for the redshift intervals $0.20 <z < 0.50$\ (filled diamonds), $0.60 < z <
1.00$\ (filled triangles; offset by +0.65~mag relative to the [0.20,0.50]
function), and $1.00 < z < 1.50$\ (open triangles; offset by +0.75~mag relative
to the [0.20,0.50] function).
}

\figcaption[]{\label{fig:10}$2000~{\rm \AA}$\ rest-frame luminosity function
in the redshift interval $0.5 < z < 0.9$\ (filled diamonds and solid line),
computed from the \uprime\ sample, compared with the luminosity function
constructed from the local rest-frame UV-selected sample of Treyer et al.\
(1998) (dashed line) for $\hno = 65~{\rm km\ s^{-1}\ Mpc^{-1}}$.  The latter
has been renormalized upward by a factor of 1.7 to match the $z = 0.7$\
luminosity function.  }

\figcaption[]{\label{fig:11}As in Fig.~10 for the redshift range $1.0 < z <
1.5$, constructed from the $B$-selected sample.  Here the local UV
luminosity function of Treyer et al.\ (1998) has been 
multiplied by a factor of 2.8 to match the $z = 1.25$\ function. 
}

\figcaption[]{\label{fig:12}Incompleteness-corrected rest-frame
$2000~{\rm\AA}$\ luminosity densities to $M_{AB} = -16$\ in the redshift
intervals $0.5 < z < 0.9$\ and $1.0 < z < 1.5$\ (filled squares), constructed
from the luminosity functions of Figs. 10 and 11, compared with the measured
$2000~{\rm \AA}$\ luminosity density from the local Treyer et al.\ (1998)
sample (filled diamond), for $\hno = 65~{\rm km\ s^{-1}\ Mpc^{-1}}$\ and $\qno
= 0.5$.  This number has been computed for the range $M_{AB} < -16$\ only.
Thin horizontal lines show the redshift ranges.  For the present data, the
open squares show the densities computed with the minimal and maximal
functions (see text and Figs.~5--7), joined by thin vertical lines.
Statistical and systematic errors (Appendix) are shown by thick vertical
lines.  The solid line is a $(1 + z)^{1.5}$\ evolution line, while the dashed
line is a $(1 + z)^4$\ evolution, normalized to the $z = 0.7$\ point.  }

\figcaption[]{\label{fig:13}Incompleteness-corrected rest-frame
$2800~{\rm\AA}$\ luminosity density as a function of redshift.  Filled squares
show the densities from the present work for the redshift intervals $0.2 < z < 
0.5$\ (constructed from the \uprime\ sample), $0.5 < z < 0.75$\ (from the $B$\
sample), $0.75 < z < 1$\ (from the $V$\ sample) and $1.0 < z < 1.5$\ (also
from the $V$\ sample).  The filled diamond is the density constructed from the
Treyer et al.\ (1998) sample corrected to $2800~{\rm\AA}$, the open diamonds
are Lilly et al.'s (1996) ``LF-estimated'' total 
$2800~{\rm\AA}$\ densities from the CFRS survey, and the open triangles show the
densities derived from Connolly et al.\ (1997) (upright symbols) and 
Sawicki et al.\ (1997) (inverted symbols), scaled downwards by 20\% as
discussed in the text.  $\hno = 65~{\rm km\ s^{-1}\ Mpc^{-1}}$\ and $\qno =
0.5$.  The dashed line shows a $(1 + z)^4$\ evolution law, similar to
the $(1 + z)^{3.9\pm0.75}$\ law derived in Lilly et al.\ (1996).  
Thin horizontal
lines show the redshift intervals in all cases.  For the present data, open
squares show the densities constructed from the minimal and maximal functions
(see text and Figs. 5--7), connected by thin vertical lines.  Thick vertical
lines show the statistical and systematic errors (Appendix).
}

\figcaption[]{\label{fig:14}$8000~{\rm\AA}$\ rest-frame luminosity functions
in the redshift intervals $0.2 < z < 0.5$\ (filled diamonds and dotted line),
constructed from the $I$-selected sample, and $0.7 < z < 1.3$\ (filled squares
and solid line), constructed from the $K$-selected sample.  $\pm 1~\sigma$\
errors are shown at each point.  
}

\figcaption[]{\label{fig:A1}$V$-band number counts (galaxies plus stars) in the
HDF proper (small squares) and in the HDF flanking fields region (large
symbols).  The dashed line is a power-law fit to the bright-end data.
}

\figcaption[]{\label{fig:A2}$B$-band galaxy counts from the HDF proper 
(Williams et al.\ 1996; small squares) compared with the average $B$-band
galaxy counts in the regions of the HDF and SSA22 used in the present work
(large symbols).  Spectroscopically confirmed stars have been removed in
both cases.
}

\figcaption[]{\label{fig:A3}Comparison of sky surface brightnesses of galaxies
in the HDF proper (filled symbols and solid line) with those in the much
larger flanking fields.  The integration excludes known stars and is over the
$AB$\ magnitude range 21 -- 24.25 in the red, 21 -- 24.5 in the visual, and 21
-- 24.75 in the blue.  The solid line shows the total sky surface brightness
for the HDF proper for comparison with the large open squares, which show this
for the flanking fields.  The smaller squares are for known $z < 1$\ galaxies
and the smaller diamonds for $z < 0.5$\ galaxies.  The dashed line shows the
possible $z > 1$\ light for this magnitude range.
}

\figcaption[]{\label{fig:A4}Rest-frame $2500~{\rm\AA}$\ luminosity functions
in the redshift range $0.6 < z < 1.0$\ constructed from subsamples of the
$B$-selected sample: the HDF plus SSA22 data (filled squares and dashed line)
and the SSA13 plus SSA4 data (open squares and solid line).  Error bars are
$\pm 1~\sigma$.  }

\newpage

\begin{figure}
\figurenum{1}
\plotone{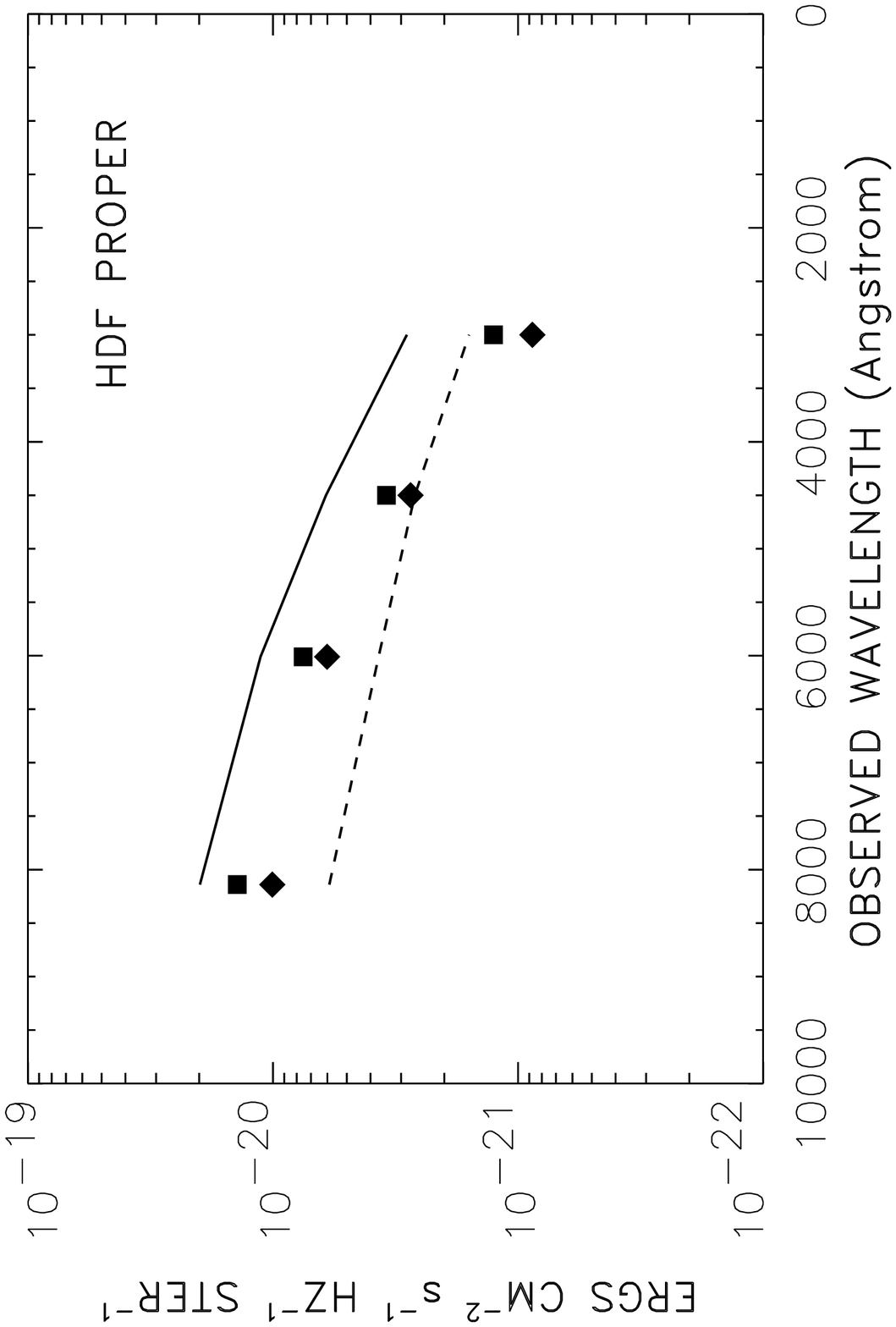}
\caption{}
\end{figure}

\begin{figure}
\figurenum{2}
\plotone{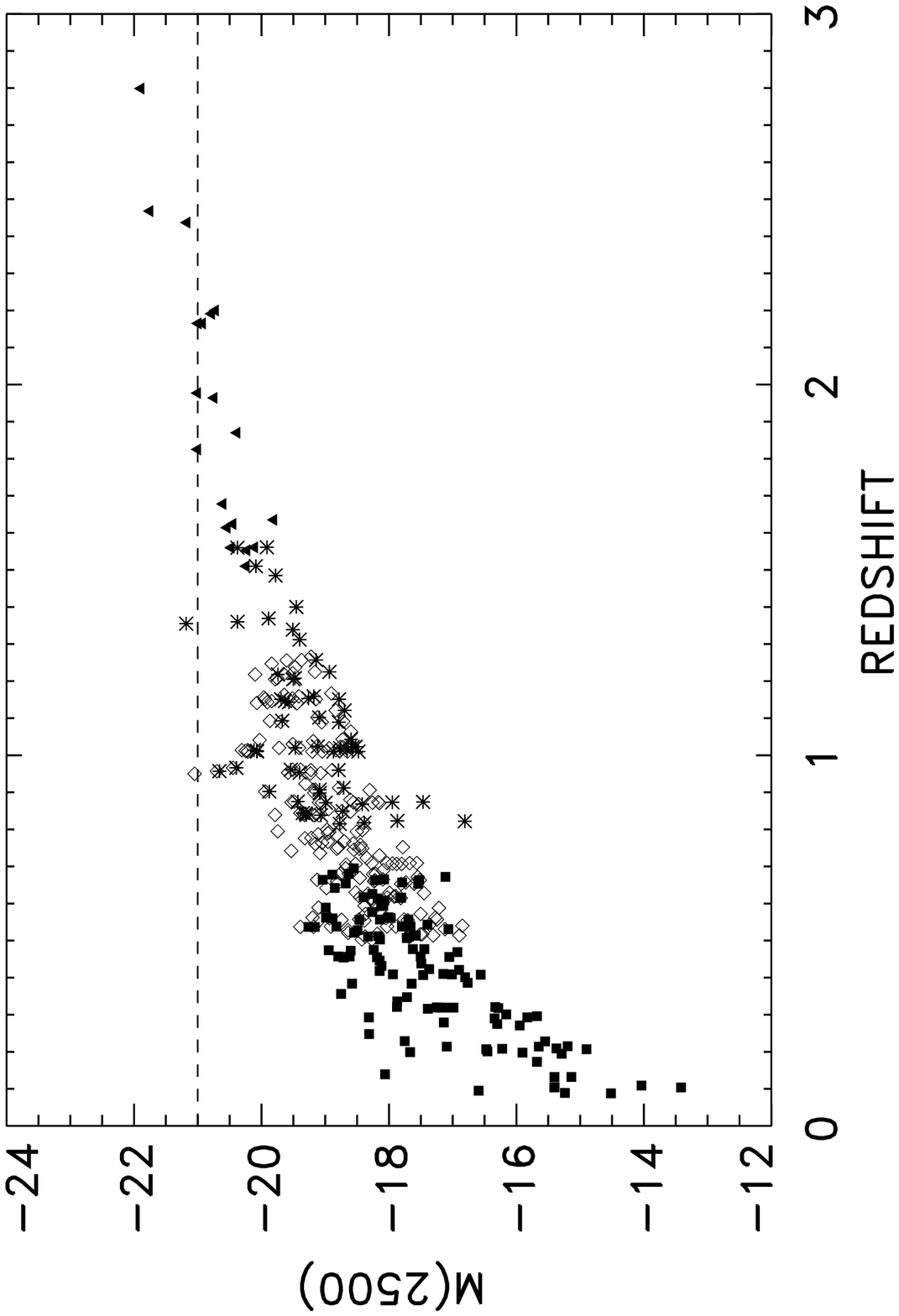}
\caption{}
\end{figure}

\begin{figure}
\figurenum{3}
\plotone{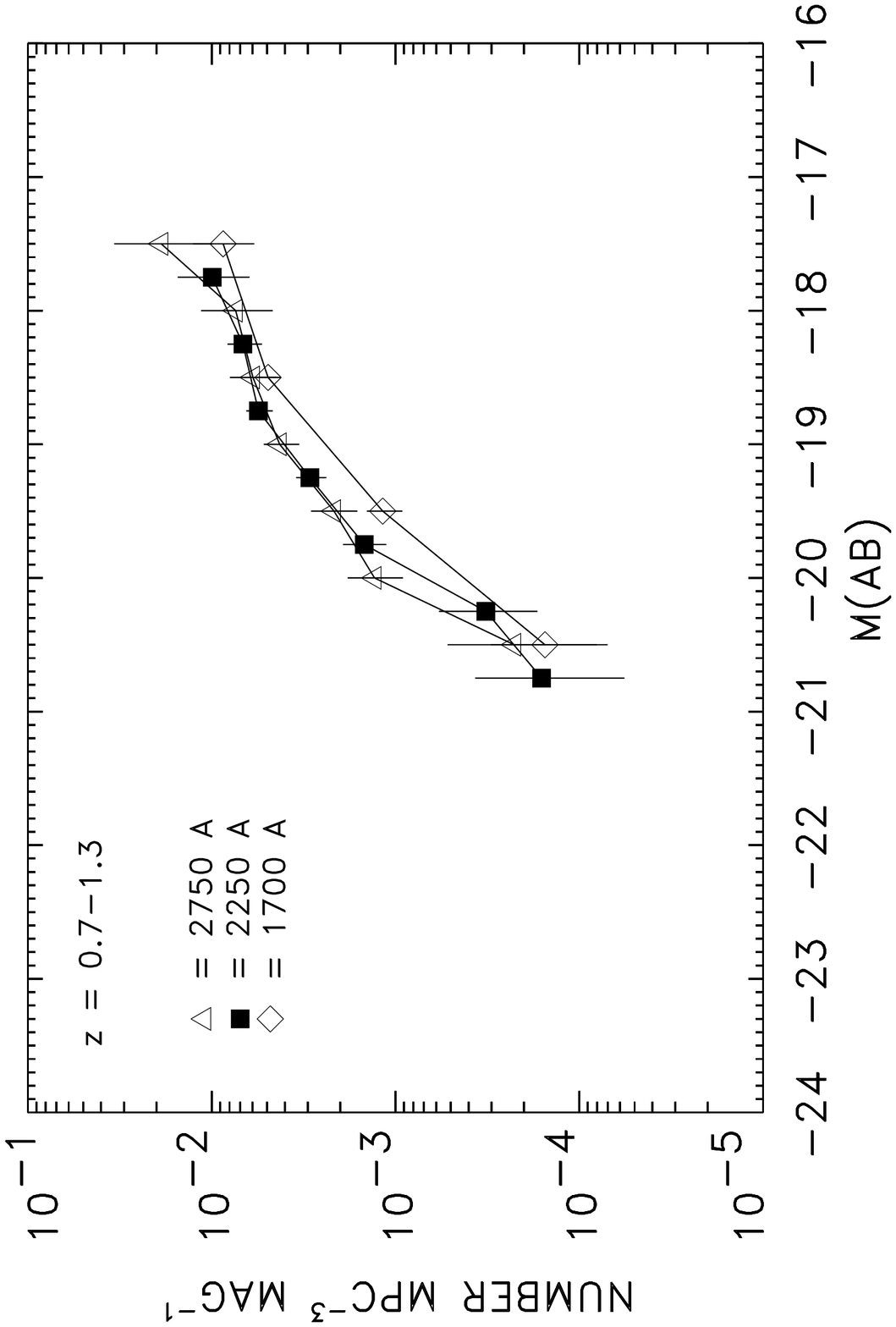}
\caption{}
\end{figure}

\begin{figure}
\figurenum{4}
\plotone{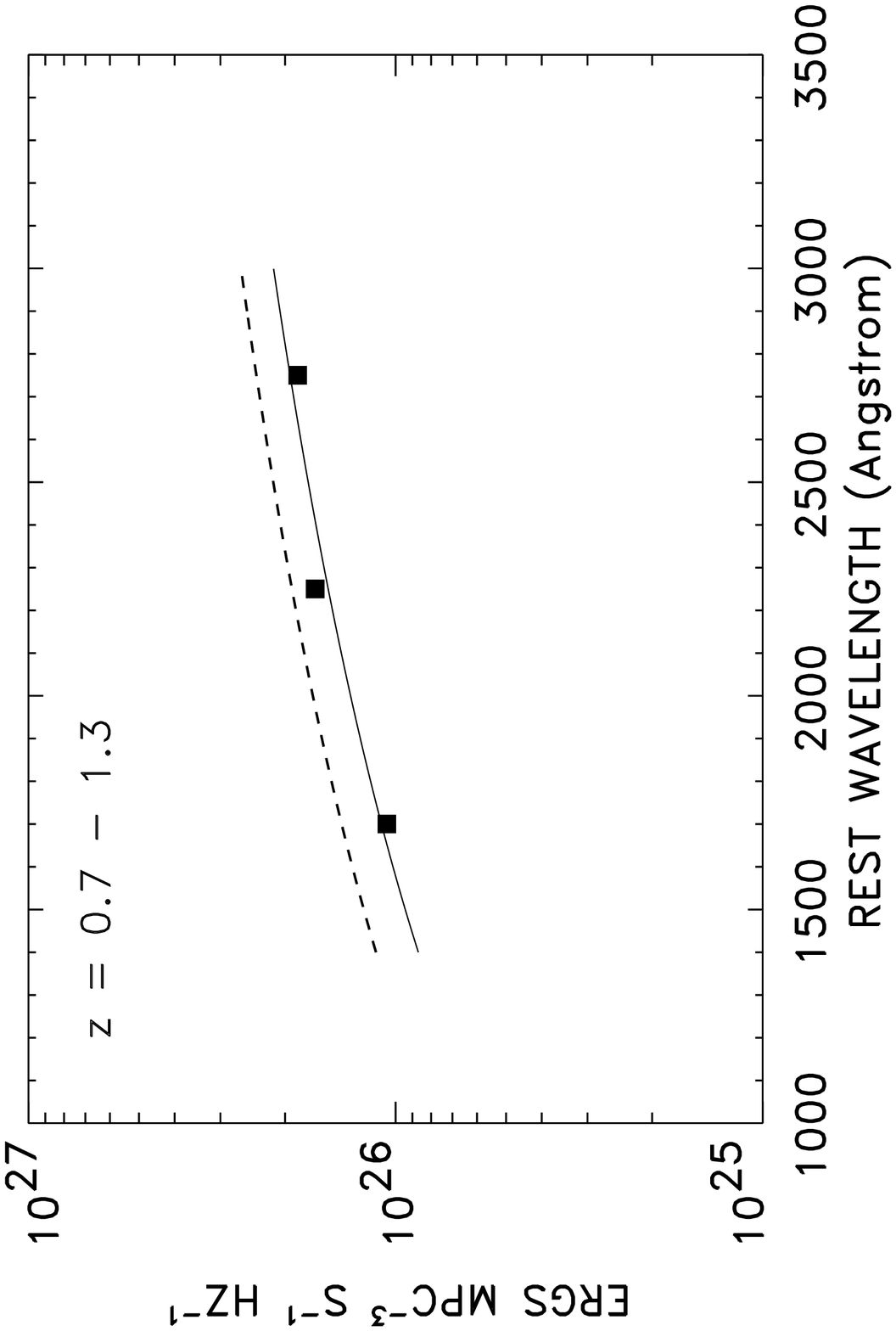}
\caption{}
\end{figure}

\begin{figure}
\figurenum{5}
\plotone{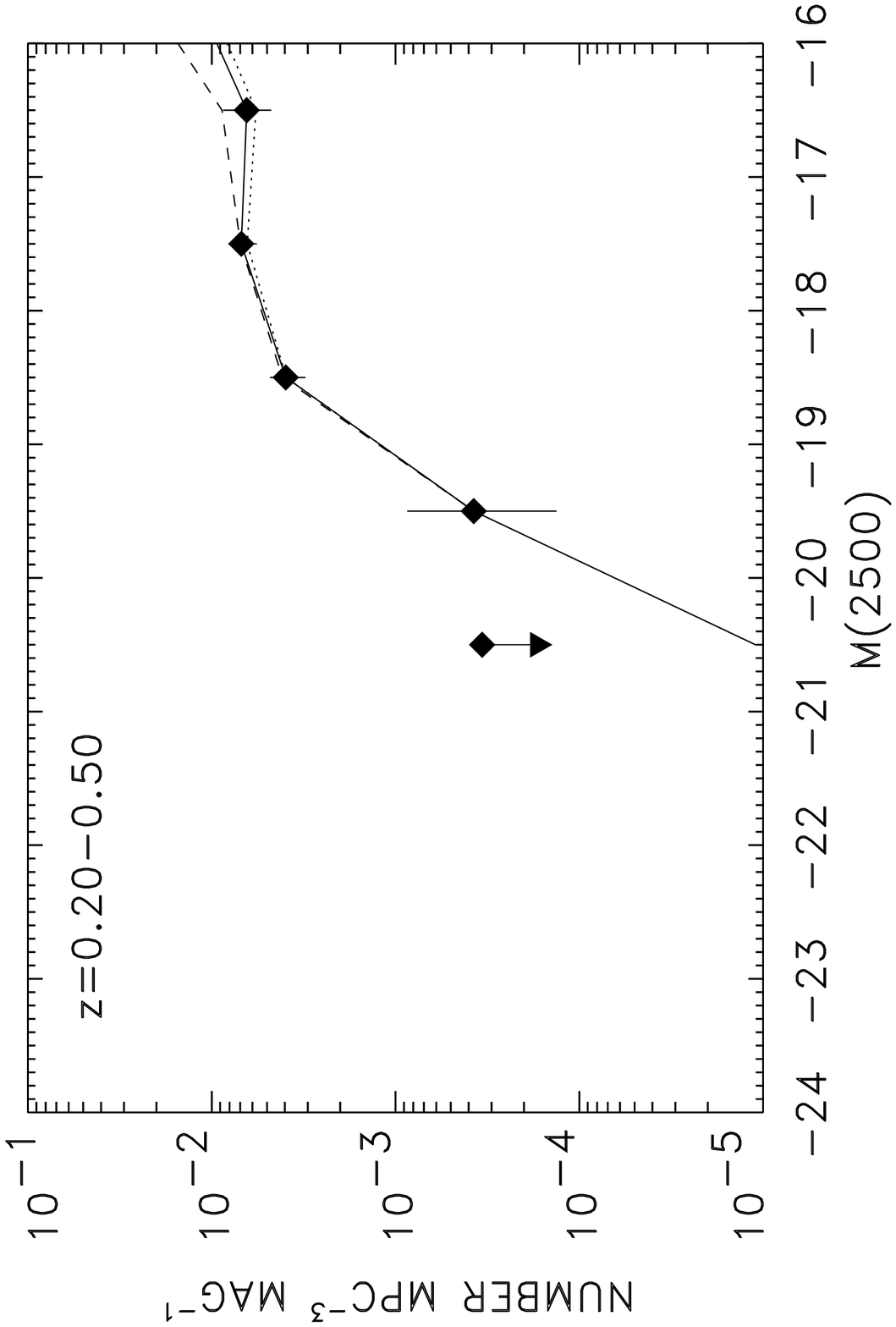}
\caption{}
\end{figure}

\begin{figure}
\figurenum{6}
\plotone{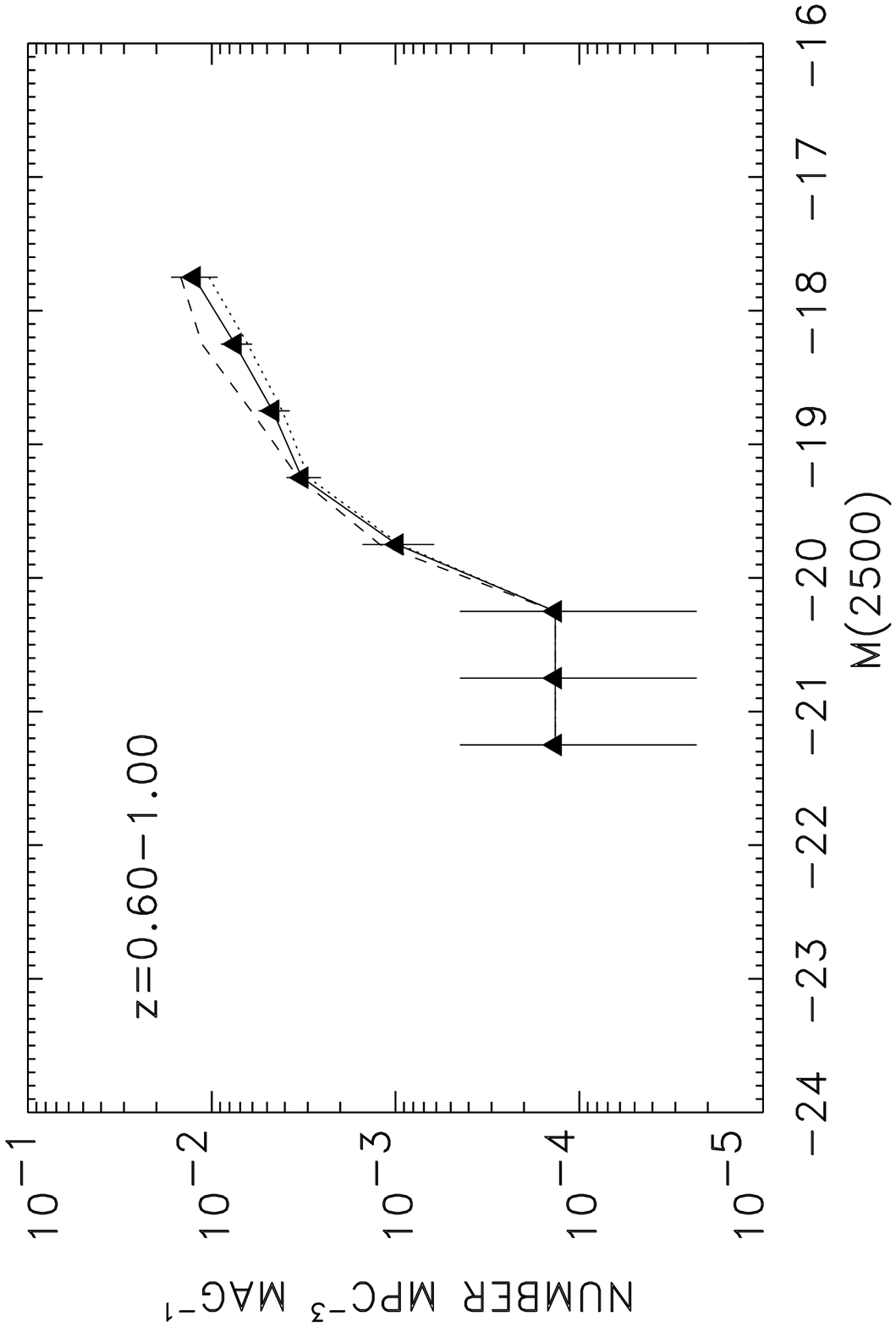}
\caption{}
\end{figure}

\begin{figure}
\figurenum{7}
\plotone{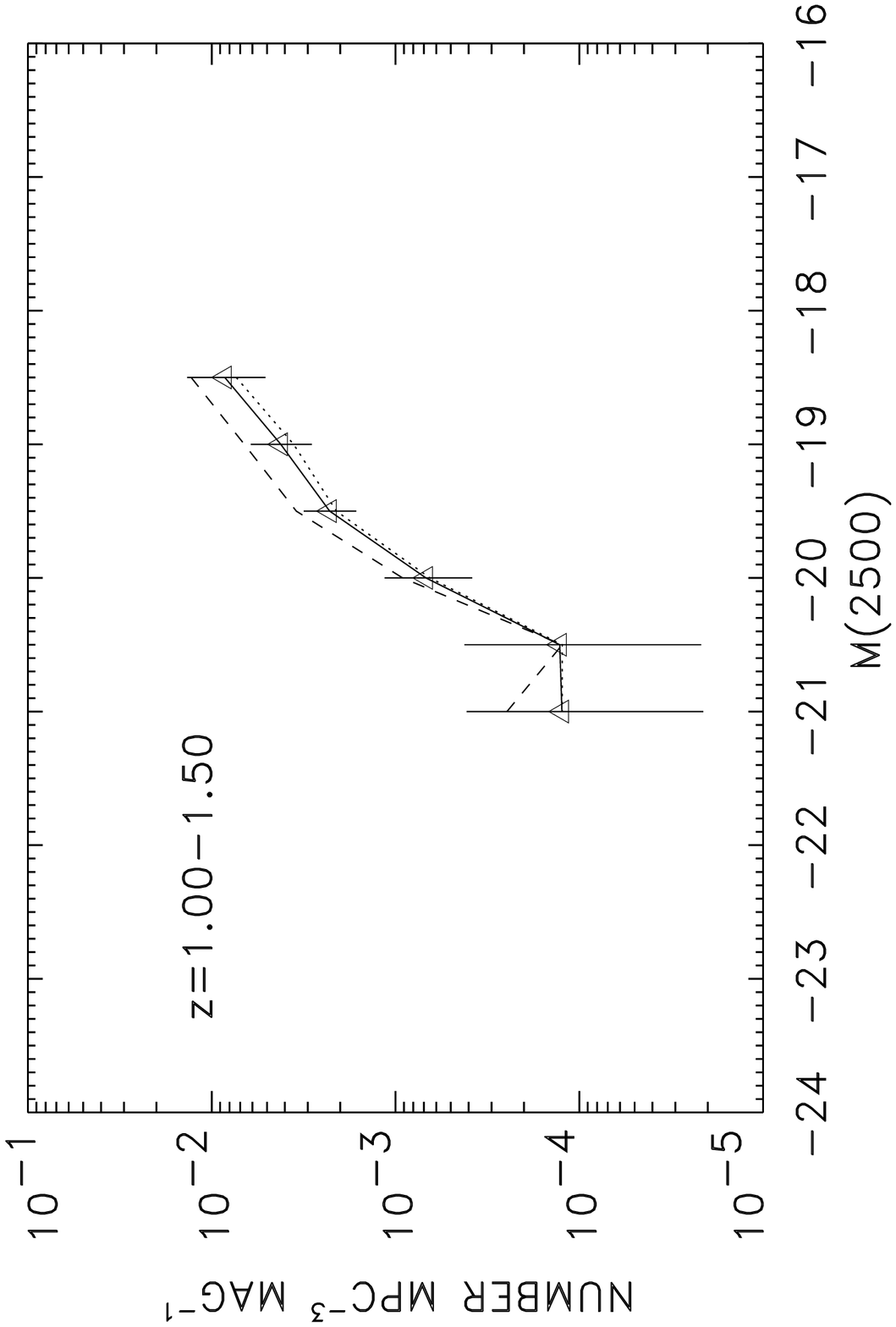}
\caption{}
\end{figure}

\begin{figure}
\figurenum{8}
\plotone{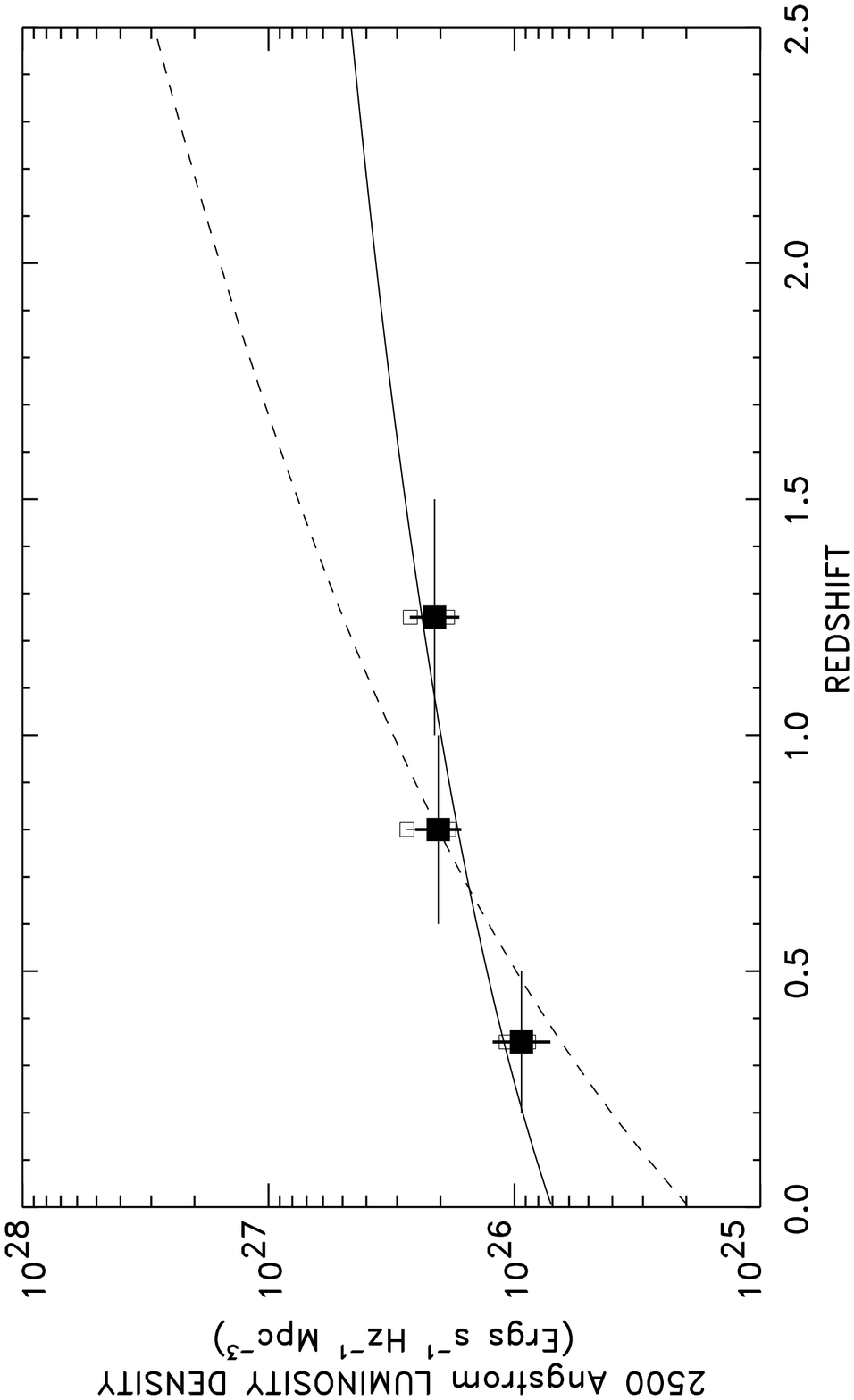}
\caption{}
\end{figure}

\begin{figure}
\figurenum{9}
\plotone{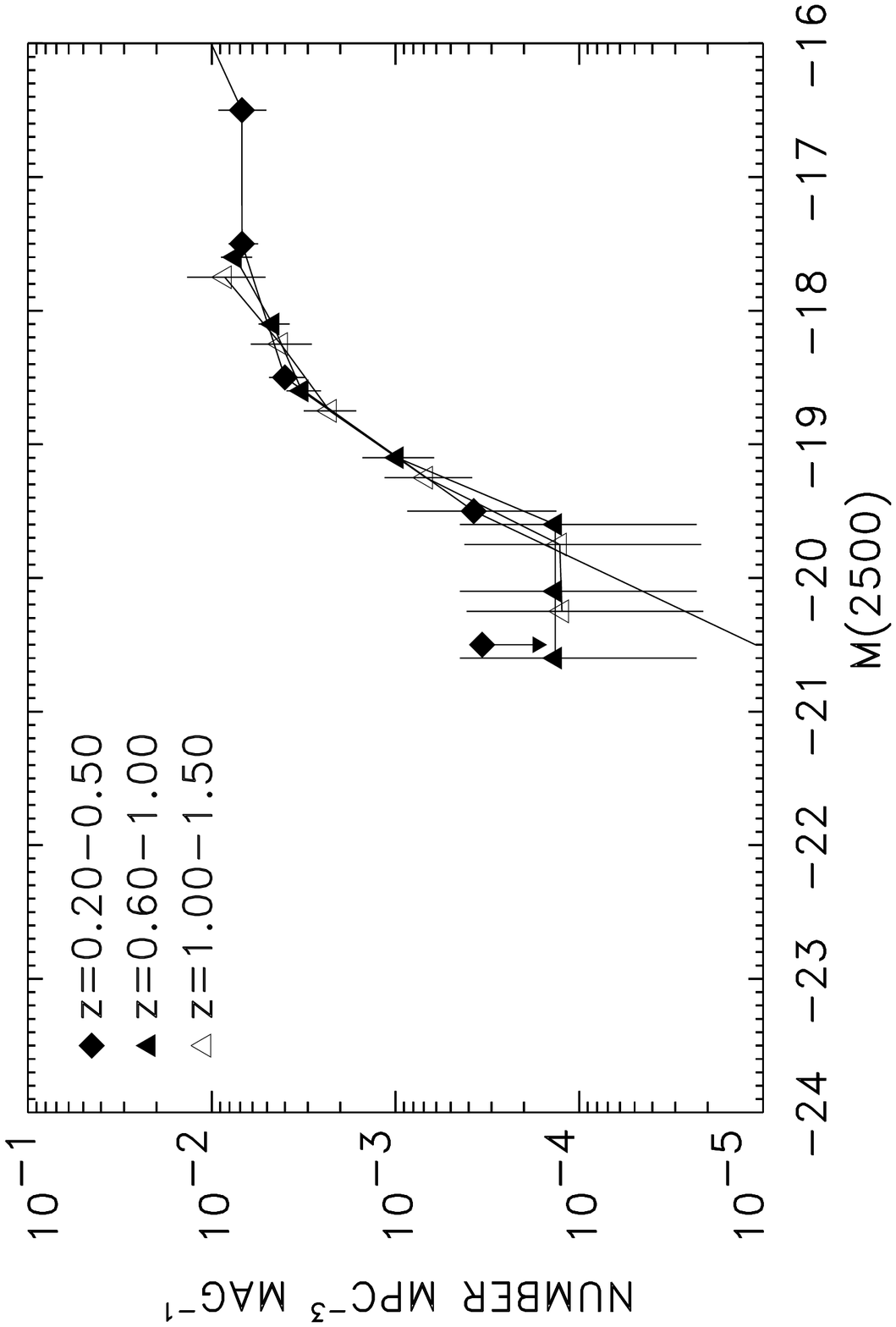}
\caption{}
\end{figure}

\begin{figure}
\figurenum{10}
\plotone{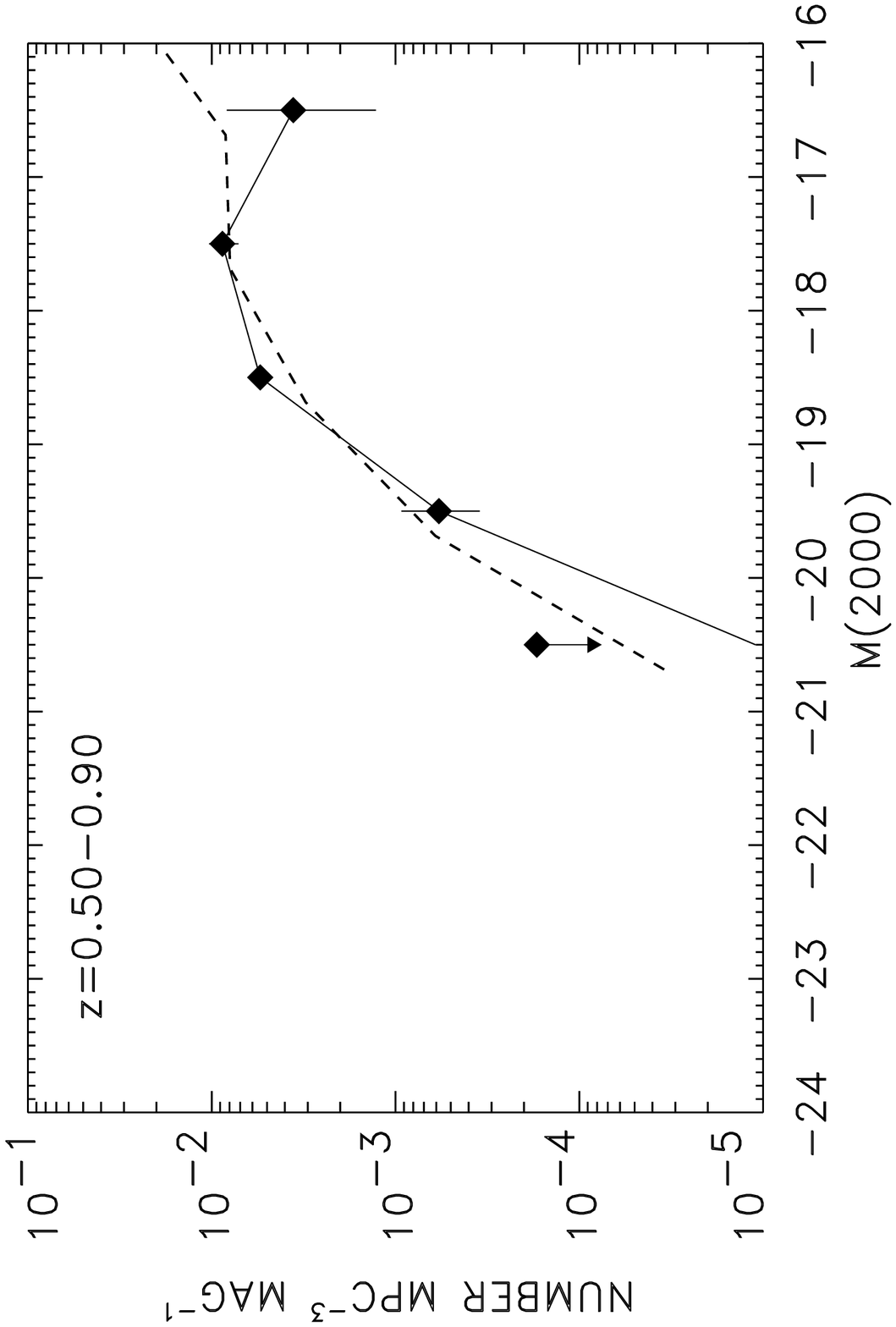}
\caption{}
\end{figure}

\begin{figure}
\figurenum{11}
\plotone{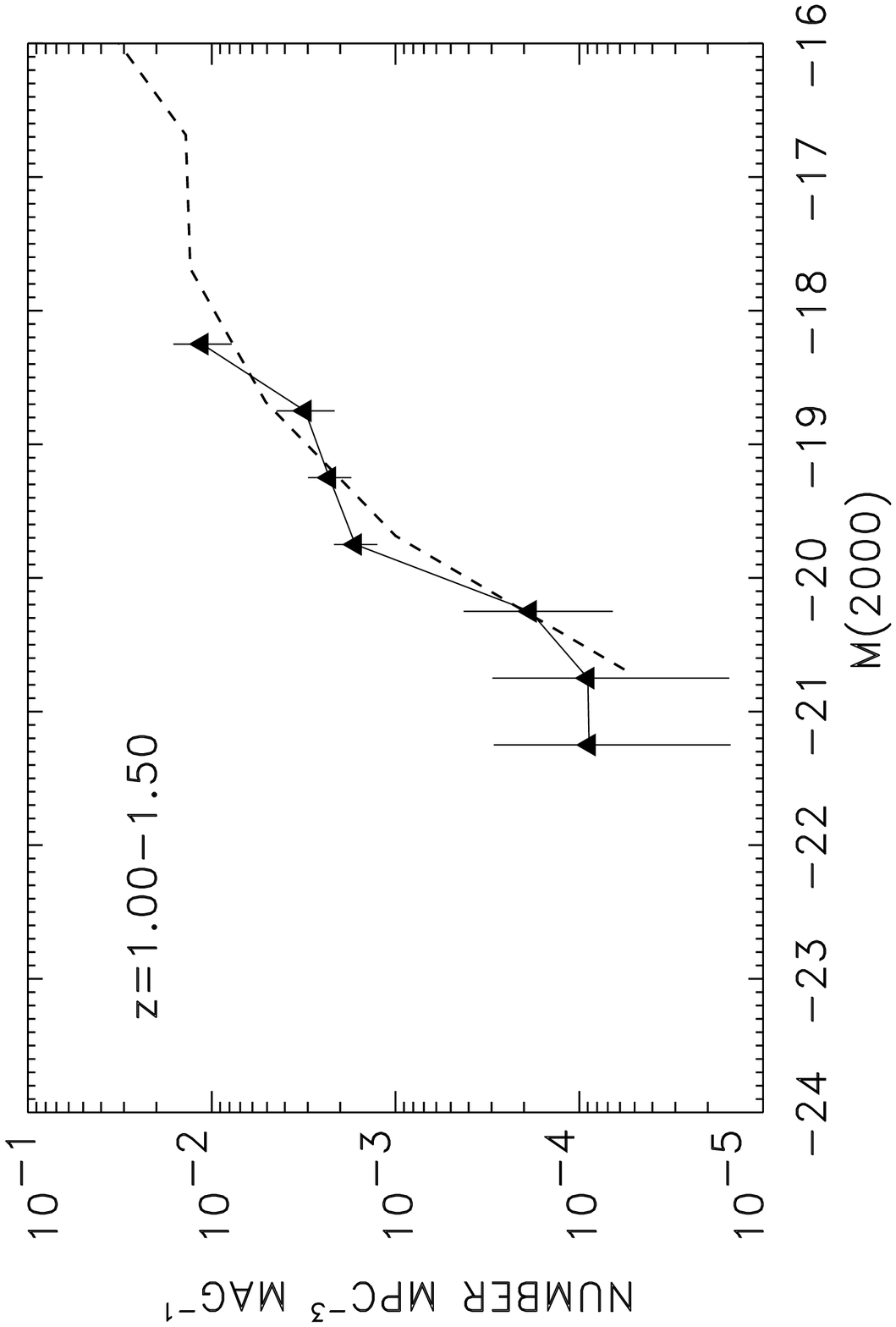}
\caption{}
\end{figure}

\begin{figure}
\figurenum{12}
\plotone{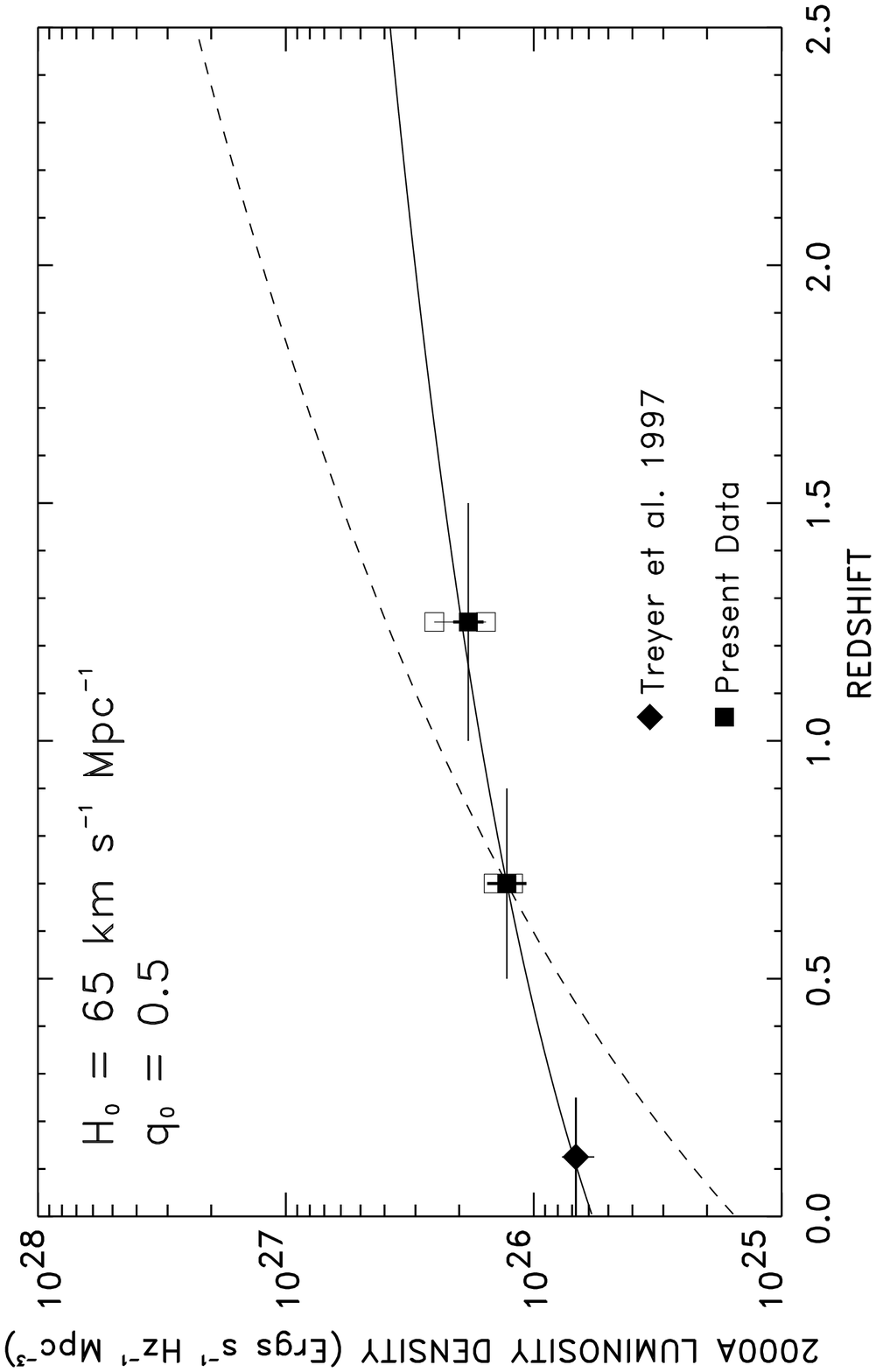}
\caption{}
\end{figure}

\begin{figure}
\figurenum{13}
\plotone{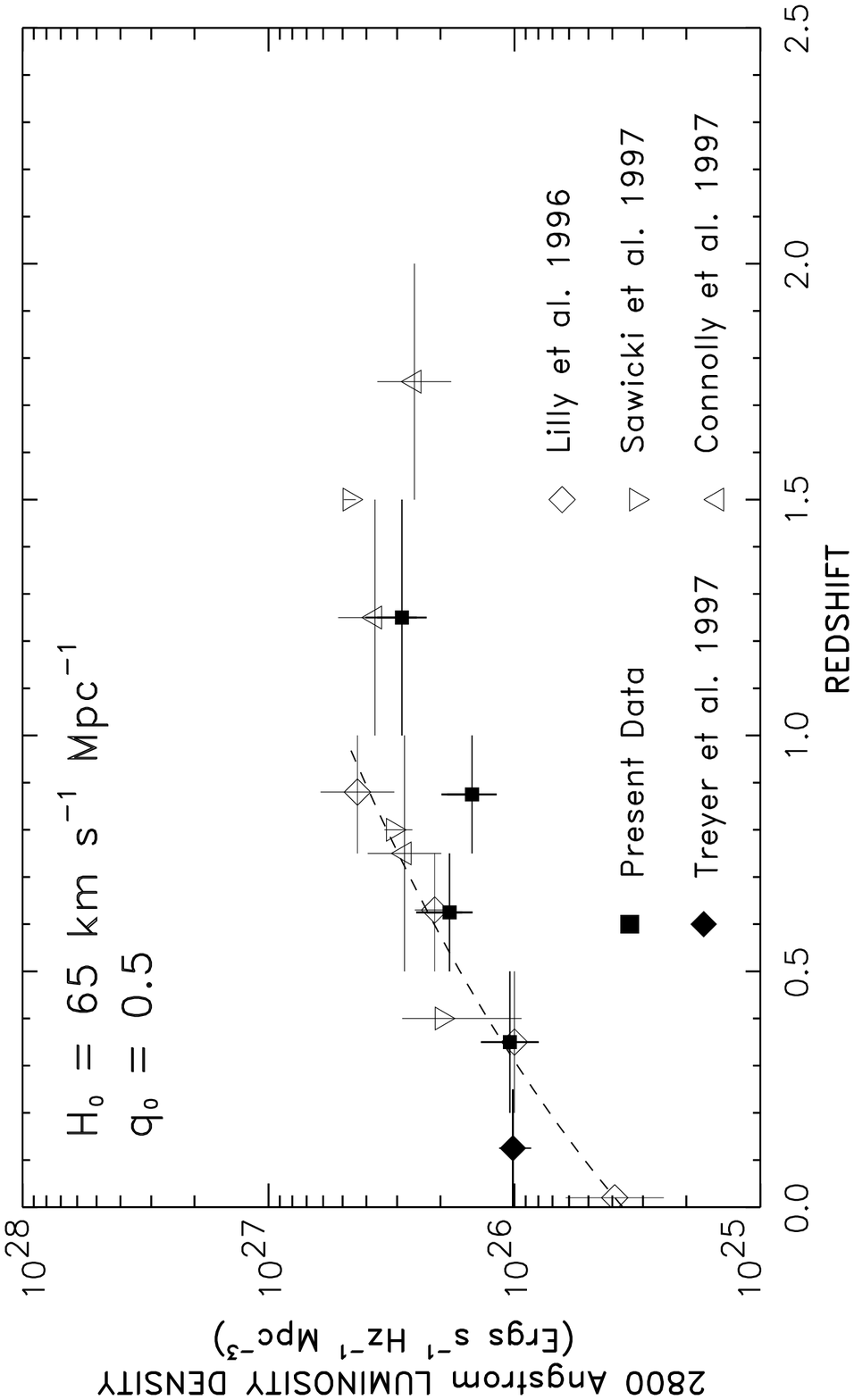}
\caption{}
\end{figure}

\begin{figure}
\figurenum{14}
\plotone{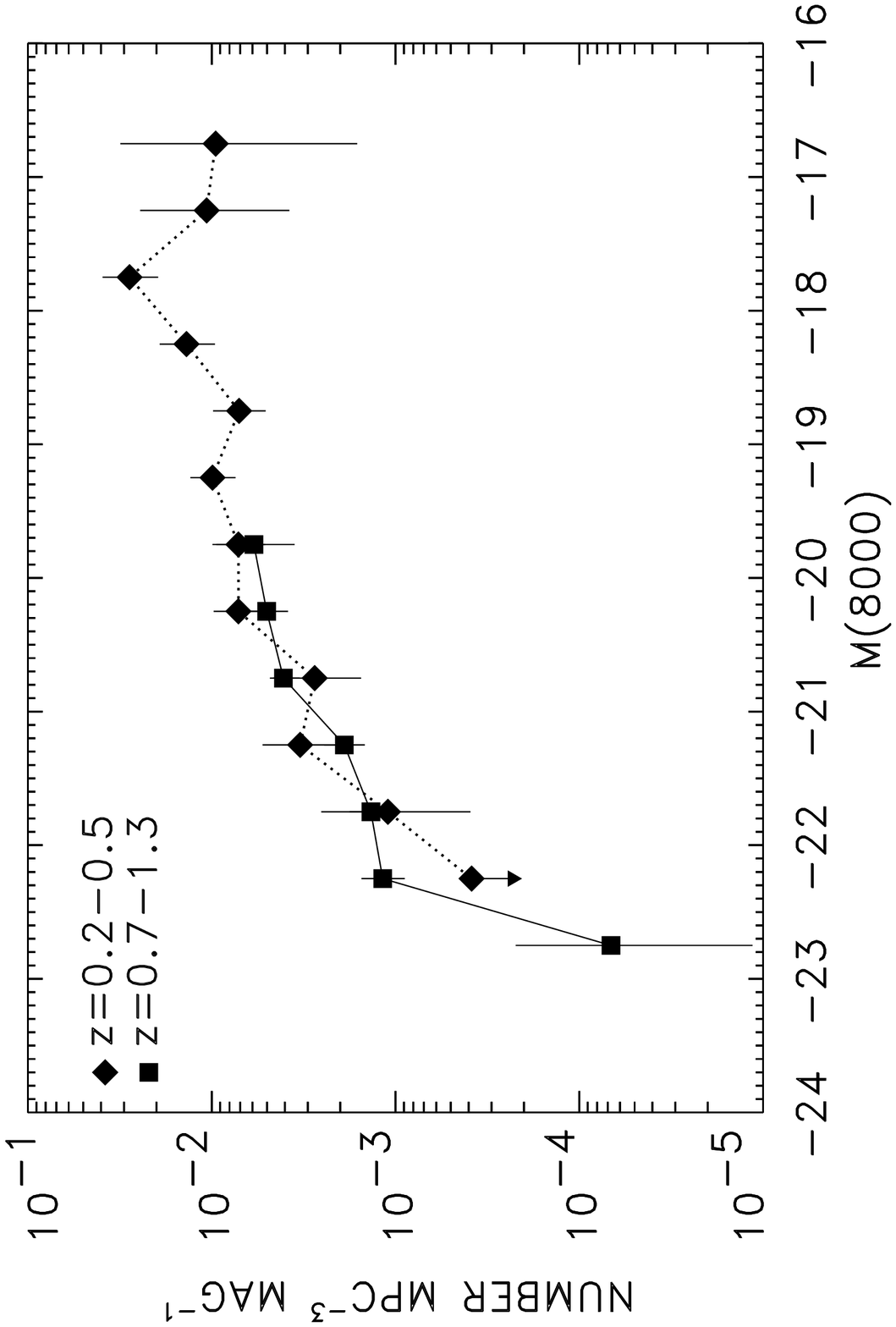}
\caption{}
\end{figure}

\begin{figure}
\figurenum{15}
\plotone{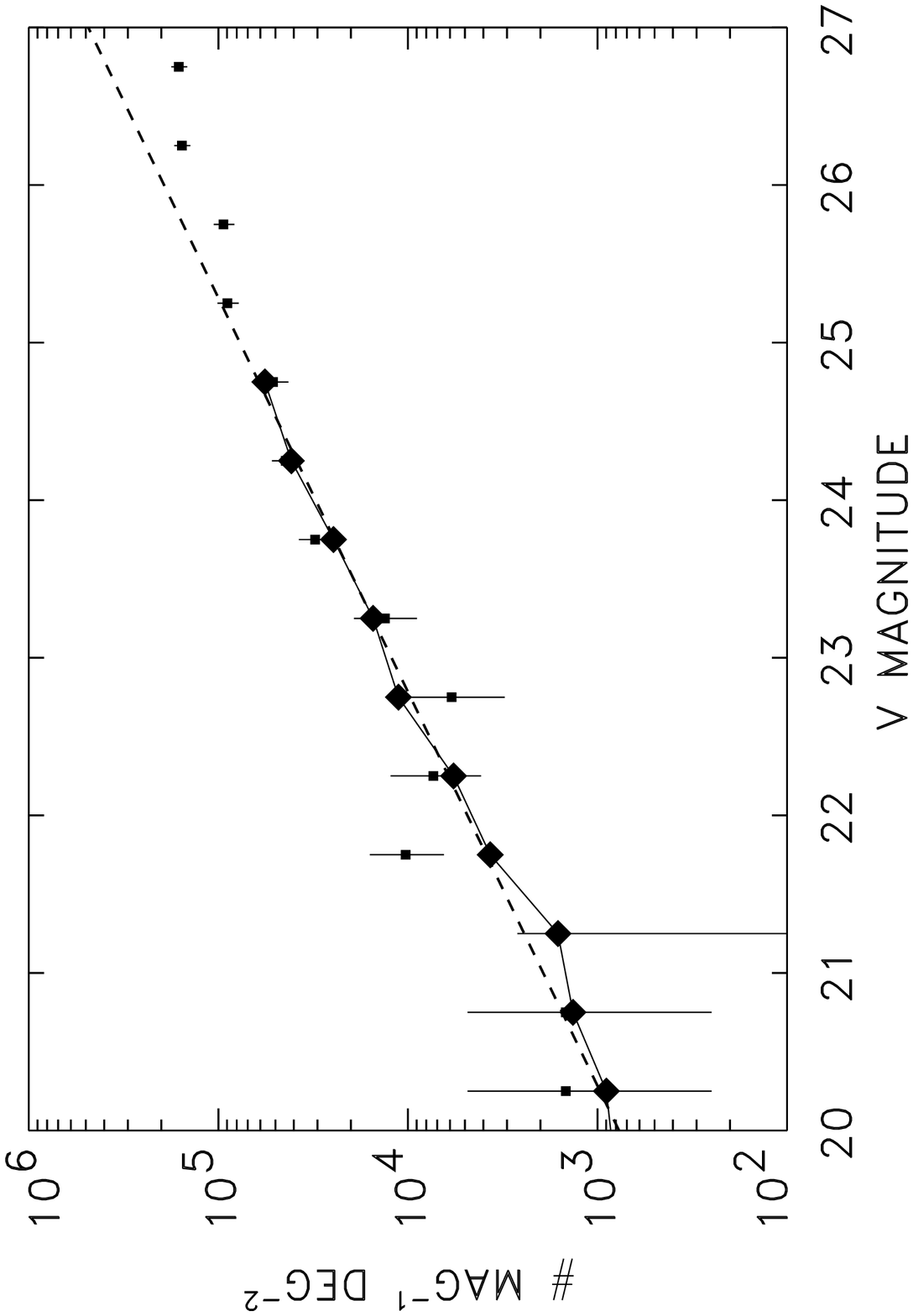}
\caption{}
\end{figure}

\begin{figure}
\figurenum{16}
\plotone{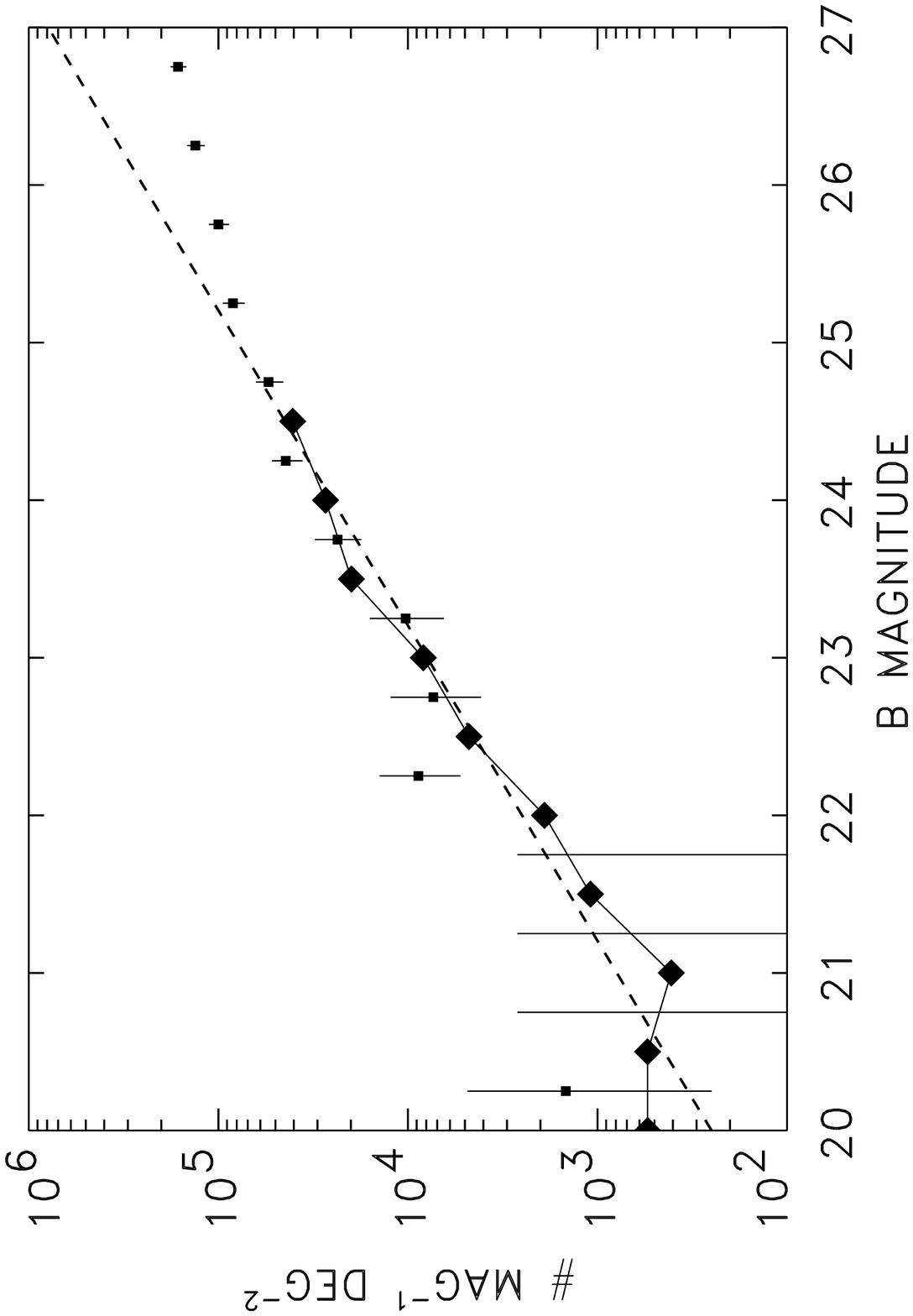}
\caption{}
\end{figure}

\begin{figure}
\figurenum{17}
\plotone{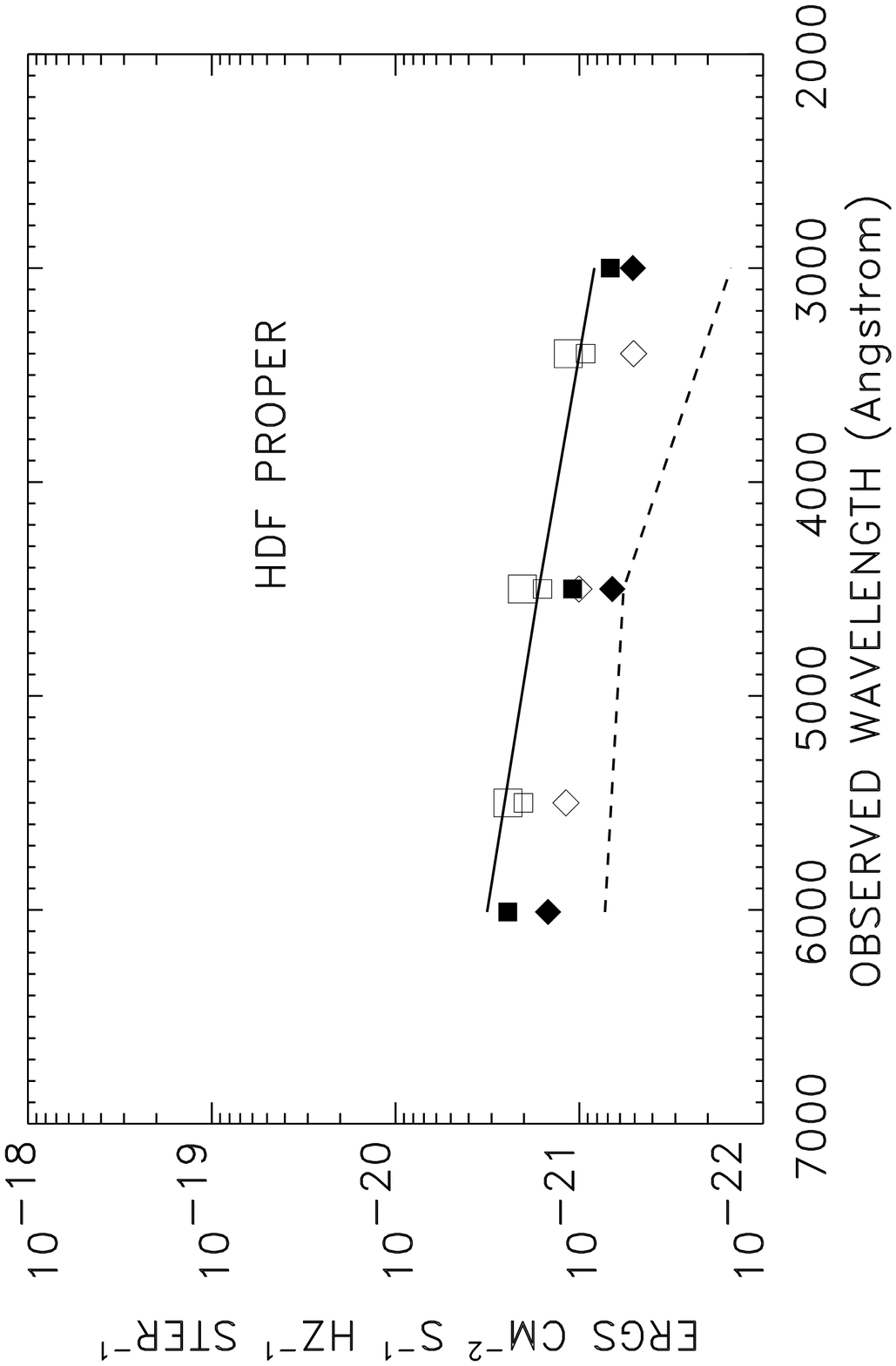}
\caption{}
\end{figure}

\begin{figure}
\figurenum{18}
\plotone{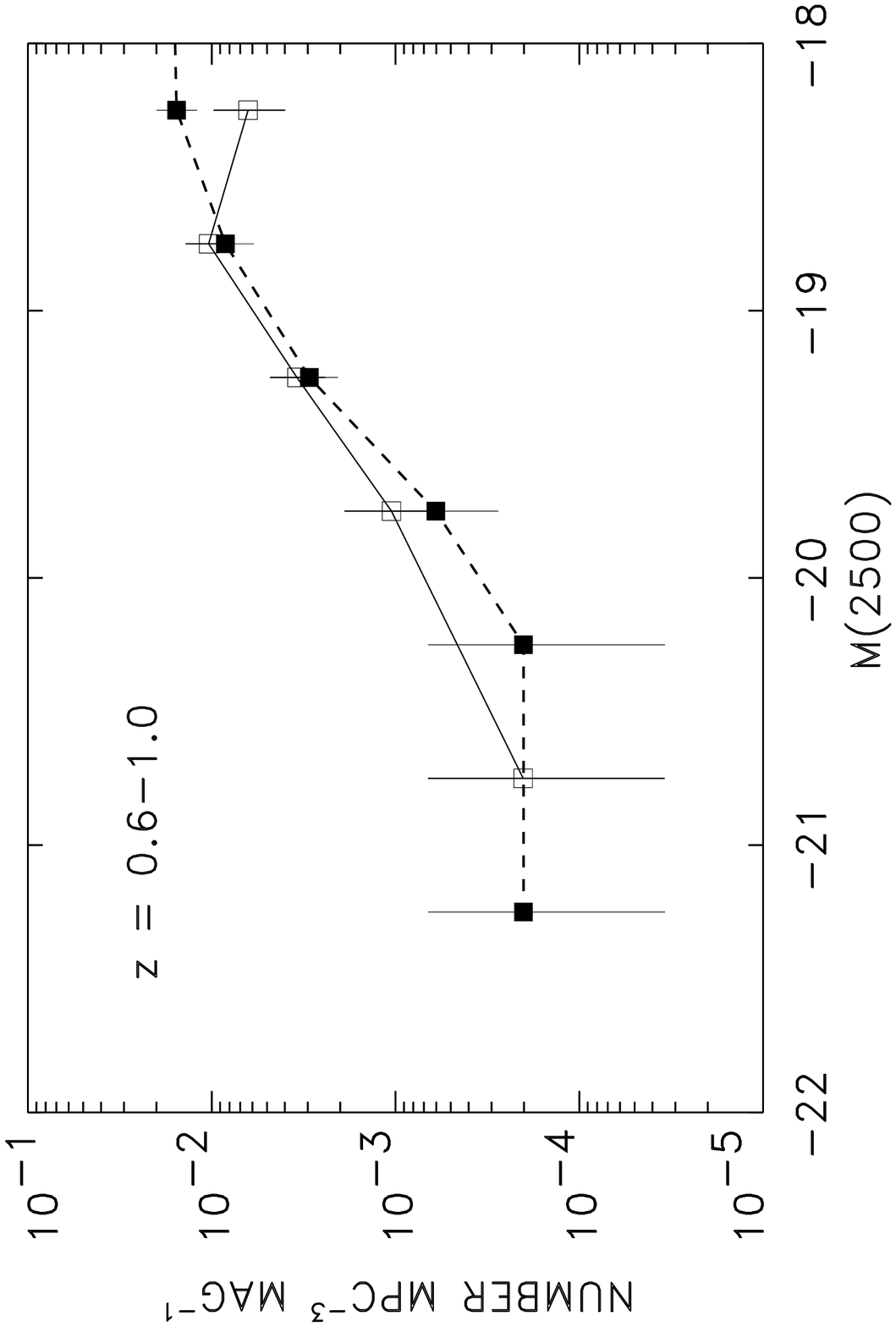}
\caption{}
\end{figure}


\newpage

\begin{references}

\reference{barg} Barger, A.J., Cowie, L.L., Trentham, N., Fulton, E., 
Hu, E.M., Songaila, A., Hall, D.\ 1999, \aj 117, 102

\reference{cohen} Cohen, J.G., Cowie, L.L., Hogg, D.W., Songaila, A.,
Blandford, R., Hu, E.M., Shopbell, P.\ 1996, \apj, 471, L5

\reference{con} Connolly, A.J., Szalay, A.S., Dickinson, M., SubbaRao, M.U.,
Brunner, R.J.\ 1997, \apj, 486, L11

\reference{cowie88} Cowie, L. L.\ 1988, in The post-recombination universe,
Proceedings of the NATO Advanced Study Institute, Cambridge, UK,
(Dordrecht: Kluwer), 1

\reference{cowie94} Cowie, L.L., Gardner, J.P., Hu, E.M., Songaila, A.,
Hodapp, K.-W., Wainscoat, R.J.\ 1994, \apj, 434, 114.

\reference{cowie96} Cowie, L. L., Songaila, A., Hu, E. M., \& Cohen,
J. G. 1996, AJ 112, 839.

\reference{ellis96} Ellis, R.S., Colless, M., Broadhurst, T., Heyl, J.,
Glazebrook, K.\ 1996, \mnras, 280, 235

\reference{felten} Felten, J.E.\ 1977, \aj 82, 861

\reference{hodapp} Hodapp, K.-W., et al.\ 1996, New Astronomy 1, 177

\reference{lcg} Lilly, S.J., Cowie, L.L., Gardner, J.P. 1991, \apj, 369, L79

\reference{lilly96} Lilly, S.J., LeF\`evre, O., Hammer, F., Crampton, D.\
1996, \apj, 460, L1

\reference{love} Loveday, J., Peterson, B.A., Efstathiou, G., Maddox, S.J.\
1992, \apj, 390, 338

\reference{low} Lowenthal, J., et al.\ 1997, \apj, 481, 673

\reference{oke} Oke, B., et al.\ 1995, \pasp, 107, 375

\reference{pasc98} Pascarelle, S. M., Lanzetta, K. M. \& Fernandez-Soto,
A. 1998,  ApJ 508, L1.

\reference{phillips} Phillips, A.C., et al.\ 1997, \apj, 489, 543

\reference{sawicki} Sawicki, M. J., Lin, H. \& Yee, H. K. C. 1997, AJ 113, 1.

\reference{song90} Songaila, A., Cowie, L.L., Lilly, S.J.\ 1990, \apj, 348, 371 

\reference{song97} Songaila, A.\ 1997, 
http://www.ifa.hawaii.edu/~cowie/tts/tts.html

\reference{steidel96} Steidel, C.C., Giavalisco, M., Dickinson, M., 
Adelberger, K.\ 1996, \aj, 112, 352

\reference{tresse} Tresse, L., Maddox, S.J.\ 1998, \apj, 495, 691

\reference{treyer} Treyer, M.A., Ellis, R.S., Milliard, B., Donas, J.,
Bridges, T.J.\ 1998, \mnras, 300, 303

\reference{will96} Williams, R., et al.\ 1996, \aj 112, 1335

\end{references}
\end{document}